\documentclass[prr,reprint,superscriptaddress,altaffillsymbol,amsmath,nofootinbib,longbibliography]{revtex4-1}

\pdfoutput=1
\usepackage{ifthen}
\usepackage{booktabs} 
\usepackage{natbib}

\usepackage{wasysym}
\usepackage{slashed} 
\usepackage{bbm}
\usepackage{mathrsfs}
\usepackage{amssymb}
\usepackage{dsfont}
\usepackage[export]{adjustbox}

\usepackage{color}

\definecolor{OurBlue}{rgb}{0.384,0.616,0.784}
\definecolor{OurRed}{rgb}{0.878,0.388,0.212}

\usepackage[
colorlinks=true, 
linkcolor=OurBlue,
citecolor=OurRed,
urlcolor=OurBlue,
]{hyperref}

\allowdisplaybreaks

\usepackage{amsmath, amssymb, dsfont, booktabs, tabularx} 

\newcommand{\beqra}{\begin{eqnarray}}
\newcommand{\eeqra}{\end{eqnarray}}
\newcommand{\beq}{\begin{equation}}
\newcommand{\eeq}{\end{equation}}

\newcommand{\LL}{\mathcal{L}}
\newcommand{\OO}{\mathcal{O}}

\def\vect#1{\boldsymbol{#1}}

\def\ve{v_\text{e}}
\def\vve{\vect{v}_\text{e}}
\def\vesc{v_\text{esc}}

\renewcommand{\epsilon}{\varepsilon}

\renewcommand{\bar}{\overline}

\newcommand{\ket}[1]{\left| #1 \right\rangle}
\newcommand{\bra}[1]{\left\langle #1 \right|}

\usepackage[columnwise]{lineno} 

\begin{document}

\title{\boldmath Chiral phonons in metal--organic frameworks as quantum sensors \\ for the direct detection of dark matter}

\author{Marek Matas}
\email[marek.matas@fjfi.cvut.cz]{}
\affiliation{Faculty of Nuclear Sciences and Physical Engineering, Czech Technical University in Prague, Czech Republic}

\author{Filip Krizek}
\affiliation{Institute of Physics, Czech Academy of Sciences, Czech Republic}

\author{Carl P. Romao}
\email[carl.romao@cvut.cz]{}
\affiliation{Faculty of Nuclear Sciences and Physical Engineering, Czech Technical University in Prague, Czech Republic}
\affiliation{Department of Materials, ETH Zurich, Switzerland}

\begin{abstract}
We investigate a new quantum sensor for dark matter direct detection with sub-eV sensitivity, focusing on several candidate materials that potentially host chiral phonons with large magnetic moments that can be directly read out with an external magnetometer. We focus on metal--organic frameworks (MOFs) as possible candidate materials for single chiral phonon detection due to their noncentrosymmetric structure, tunability, and the ability to host these excitations in stable acoustic bands. We identify several promising candidates and compare their projected dark matter detection sensitivity for all possible interactions identified within effective field theory. We establish that the expected sensitivity does not depend heavily on the specific choice of the MOF, enabling us to tailor the final material composition to facilitate the magnetic readout. We then propose a prototype setup able to test the direct readout of a chiral phonon sensor with a surface-integrated magnetometer.
\end{abstract}

\maketitle

\section{Introduction}

Dark matter (DM), one of the long-standing mysteries of modern physics, has attracted considerable attention both from experiment and theory in recent years~\cite{Cirelli:2024ssz}. As evidence builds against explanations that require a modification of today’s laws of physics, the interest in capturing the particle-like nature of DM is increasing~\cite{Clowe:2006eq, Harvey:2015hha, vanDokkum:2018vup, van_Dokkum_2019}.

To date, no unambiguously confirmed signal has been observed. The effort to capture a direct interaction between a DM particle and a Standard Model (SM) particle in a terrestrial detector therefore proceeds along two main directions: either to increase the exposure and probe weaker interactions~\cite{XENON:2019gfn, XENON:2023cxc, DarkSide-20k:2017zyg}, or to lower the detection threshold and explore smaller DM masses. While scaling detectors up and keeping their backgrounds low is becoming increasingly difficult and costly, small-scale table-top experiments with better sensitivity can probe inaccessible regions of DM phase space, breaking world-leading limits~\cite{PhysRevLett.123.151802, qrocodile, Gao:2024irf}.

The emerging field of quantum sensing is one of the pathways to open up detection of energy depositions below that of a single semiconductor exciton. With devices such as superconducting nanowire single-photon detectors (SNSPDs) and transition edge sensors (TESs), one can, in principle, reach energy depositions of the order of $\mathcal{O}(10)$\,meV~\cite{qrocodile}, while proposed space-based gradiometers will be capable of sensing fluctuations of much lighter DM particles probing the wave-like hypothesis at a new scale~\cite{Badurina:2025xwl}.

In this work, we discuss a new quantum sensor that would allow for a non-thermal detection of individual phonons with energy depositions of $\mathcal{O}(10)$\,meV to join the hunt for DM~\cite{Romao:2023zqf}. This sensor is based on the properties of chiral phonons, which are quantized vibrational excitations in solids wherein the atoms revolve around their average positions~\cite{juraschekChiral2025}. This circular motion yields a net angular momentum, which is carried by the phonon. Chiral phonons naturally arise at arbitrary points in reciprocal space in noncentrosymmetric crystals~\cite{quartz}.

It has been reported that angular momentum-carrying phonons can have large magnetic moments, on the order of the Bohr magneton~\cite{shabalaAxial2025}. This surprising discovery has been experimentally observed through the in-field Zeeman shift of the phononic states~\cite{schaack1976observation, wuValley2018, cheng2020large, BaydinPbTe, hernandezObservation2023}, as well as by the Kerr ~\cite{luo2023large, basiniTerahertz2024} and Faraday~\cite{luo2023large, hernandezObservation2023, biggsUltrafast2025} effects in optical pump--probe experiments. Phonon magnetism in magnetic materials can be explained by hybridization between spin excitations and phonons~\cite{juraschek2022giant, luo2023large, luoEvidence2023, chaudharyGiant2024}, whereas in nonmagnetic materials several mechanisms involving electron--phonon coupling have been proposed~\cite{juraschek2017dynamical, ren2021phonon, zabaloRotational2022, geilhufeDynamic2022, shabalaPhonon2024, chenGauge2025}.

The phonon magnetic moments found in some materials are large enough to be read out by sensitive magnetometers such as superconducting quantum interference devices (SQUIDs)~\cite{vasyukov2013scanning} or nitrogen vacancy (NV) centers~\cite{grinolds2013nanoscale}. Even smaller magnetic moments could be detectable if larger numbers of phonons are excited, as the magnetic moments would be additive due to their bosonic nature. This would allow us to detect individual phononic excitations directly for the first time and open up a pathway for their use as sensors, both for probing an uncharted parameter space of dark matter as well as for use as a new general type of quantum sensor~\cite{Romao:2023zqf}.

In order to construct a prototype of the device, we need to identify the most promising candidate material. The material should be non-centrosymmetric in order to be able to host intrinsically chiral phonons at an arbitrary point in reciprocal space. It should host them in the stable acoustic bands ~\cite{maris1993anharmonic}, and these phonons should possess detectable magnetic moments.

Metal-organic frameworks (MOFs) are a promising family of materials that satisfy all the conditions above~\cite{wang2012rational}. Their structure is often non-centrosymmetric, with the ligands allowing for an easy movement of the bound atoms around their equilibrium positions. This leads to a low energy of the chiral phononic excitations, allowing them to appear in the stable low-energy transverse acoustic band~\cite{Romao:2023zqf}. MOFs can possess additional degrees of freedom, such as molecular rotation, which could provide additional mechanisms for producing phonon magnetism through spin-rotation coupling \cite{huangQuantum2024, shabalaAxial2025} MOFs are often modifiable, where one can exchange the constituent atoms and tune the phonon magnetic moment, for example by introduction of electronic topology \cite{deng2021designing, ni2022emergence}. 

\begin{figure*}[ht]
    \centering
    \includegraphics[width=0.99\textwidth,trim=0cm 12cm 4cm 0cm,clip]{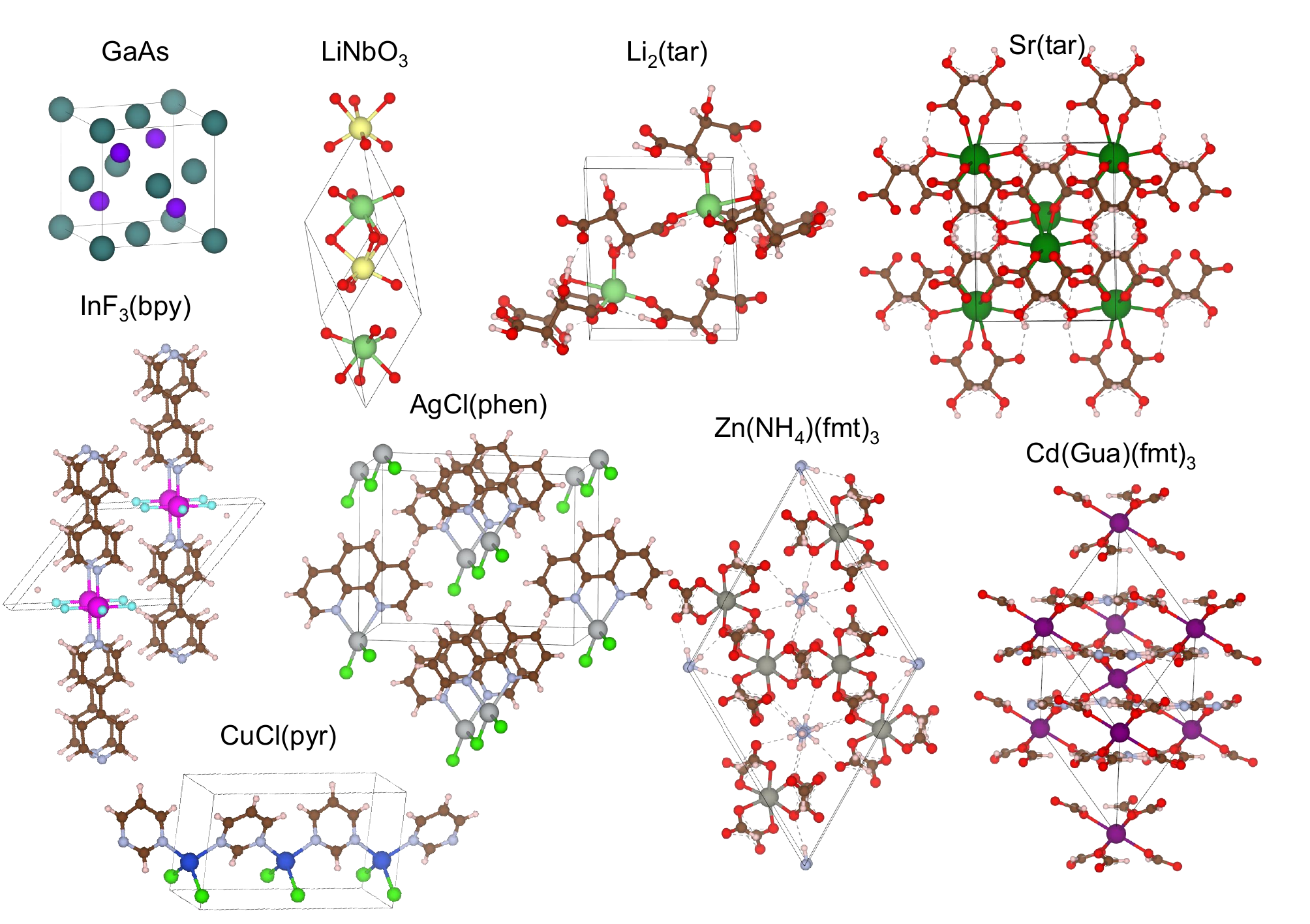}
  \caption{Crystal structures of noncentrosymmetric materials studied as candidate chiral phonon DM detector targets: GaAs (space group $F\bar{4}3m$), LiNbO$_3$ ($R3c$), Li$_2$(tartrate)  (Li$_2$(tar), $P2_1$), Sr(tartrate) (Sr(tar),  $C222_1$), InF$_3$(bipyridine) (InF$_3$(bpy), $I222$), AgCl(phenanthroline) (AgCl(phen), $C2$), Zn(NH$_4$)(formate)$_3$, (Zn(NH$_4$)(formate)$_3$, $P6_3$) Cd(guanadinium)(formate)$_3$ (Cd(Gua)(fmt)$_3$, $Cc$), and CuCl(pyrimidine) (CuCl(pyr), $Pma2$).  H atoms are shown in pink, Li atoms in pale green, C atoms in brown, N atoms in lilac, O atoms in red, F atoms in cyan, Cl atoms in bright green, Cu atoms in blue, Zn atoms in dark grey, Ga atoms in blue-green, As atoms in violet, Sr atoms in dark green, Nb atoms in yellow, Ag atoms in light grey, Cd atoms in purple, and In atoms in magenta. Unit cell vectors are marked as black lines.}
  \label{fig-crystal-structures}
\end{figure*}

Herein, we explore several candidate materials (Fig.~\ref{fig-crystal-structures}), including MOFs and conventional inorganic crystals, for the construction of a chiral phonon detector prototype. We perform \textit{ab initio} calculations of their phononic structure and determine the phonon chirality. We then simulate the interaction sensitivity of a DM detector assuming zero background and 95\% exclusion limits with the use of the PhonoDark~\cite{Trickle:2020oki} and Phonopy~\cite{TOGO20151} software packages.

\section{Phononic excitations induced by dark matter}\label{sec:dm-theory}
In this work, we follow the notation and formalism of Refs.~\cite{Coskuner:2021qxo, Trickle:2020oki, Griffin:2019mvc, Trickle:2019nya} while focusing on nonmagnetic materials only, which would allow for a clean magnetometric readout of the chiral phonon signal. We employ the non-relativistic effective field theory expansion of DM interactions, which utilizes Galilean invariance and momentum conservation to restrict the possible pathways of interaction a DM particle can have~\cite{Catena:2019gfa, Catena:2022fnk, PhysRevD.98.123003, Catena:2019hzw, Catena:2021qsr}. In the case of electronic and nuclear recoils, this theoretical framework yields fourteen independent interaction operators with a spin-1/2 DM particle~\cite{Fitzpatrick:2012ix}.

For light DM particles ($m_\chi < 10$\,MeV), which are the focus of this work, momentum transfer cannot resolve the sub-ionic degrees of freedom and probes whole atoms. Therefore, under this long-wavelength (or equivalently small momentum transfer) limit, the relevant quantities that characterize the target in a DM scattering process are the total particle numbers $\langle N_\psi\rangle$, the total spins $\langle S_\psi\rangle$, the orbital angular momenta $\langle L_\psi\rangle$, and the spin-orbit couplings $\langle L_\psi \otimes S_\psi \rangle$. For the materials considered within this work, effective operators coupling to spin, orbital angular momenta, and spin-orbit coupling are expected to be strongly suppressed with the total spin number $\langle S_\psi\rangle$ and electronic orbital angular momenta $\langle L_\psi\rangle$ averaging out.

Under such constraints, the remaining non-zero effective operators identified in Refs.~\cite{Trickle:2020oki, Fitzpatrick:2012ix} are
\begin{eqnarray}
\OO_{1}^{(\psi)}  &=& \mathds{1} \nonumber\,, \\
\OO_{5a}^{(\psi)} &=& \vect{S}_\chi \cdot \left( \frac{i\vect{q}}{m_\psi} \times \vect{v}_\chi \right) \nonumber\,, \\
\OO_{8a}^{(\psi)} &=& \vect{S}_\chi \cdot \vect{v}_\chi \nonumber\,, \\
\OO_{11}^{(\psi)} &=& \vect{S}_\chi \cdot \frac{i\vect{q}}{m_\psi}\,,
\label{eq:operators}
\end{eqnarray}
where $\vect{q} \equiv \vect{p'} - \vect{p} = \vect{k'} - \vect{k}$ is the momentum transferred from the DM particle to the target material with $\vect{p}$ ($\vect{p'}$) denoting the momentum of DM before (after) the collision and $\vect{k}$ ($\vect{k'}$) the momentum of the struck fermion. $\vect{S}_\chi$ is the DM spin operator, $m_\psi$ is the mass of the fermion, and $\mathds{1}$ is a $2\times2$ unit matrix in the spin space.

The average DM velocity $\vect{v}_\chi$ is defined with the average fermionic velocity $\vect{v}_\psi$ as
\begin{equation}
\vect{v}_\chi \equiv \frac{\vect{p'} + \vect{p}}{2m_\chi} \,,\qquad\quad
\vect{v}_\psi \equiv \frac{\vect{k'} + \vect{k}}{2m_\psi} \,,
\label{eq:two_velocities}
\end{equation}
and together they form additional independent combinations of momenta that the interaction can depend on~\cite{Trickle:2020oki}. 

For the case of electronic and nuclear recoils, these could be further reduced into a single combined velocity $\vect{v^\perp} = \vect{v}_\chi - \vect{v}_\psi$~\cite{Catena:2019gfa}, which is not possible for the collective excitations considered throughout this work. Here, we must treat the contributions of both velocities independently and thus split operators $\OO_3$, $\OO_5$, $\OO_7$, $\OO_8$, $\OO_{12}$, $\OO_{13}$, $\OO_{14}$, and $\OO_{15}$ into two parts dependent on each velocity component, denoted either with subscript $a$ (depending on $\vect{v}_\chi$) or $b$ (depending on $\vect{v}_\psi$). The $\vect{v}_\psi$-dependent part of the operator couples the incoming DM particle to the velocity of the bound fermion. This manifests itself as a coupling either to the orbital angular momenta $\langle L_\psi\rangle$, or spin-orbit couplings $\langle L_\psi \otimes S_\psi \rangle$, that are suppressed in our calculations, leaving us solely with the $\vect{v}_\chi$-components of the interaction operators~\cite{Trickle:2020oki}.

In this framework, a DM particle interacts with a fermion within the detector material, either with an electron or a nucleon ($\psi = \text{p, n, e}$), and deposits energy that causes a measurable collective excitation. Therefore, one cannot omit the leptophilic (interacting only with electrons) or hadrophilic (interacting only with protons and neutrons) dark matter models for having different interaction scales, as they independently contribute to the interaction.

The formalism of phononic interactions builds upon Fermi's golden rule in the form
\begin{equation}
\Gamma(\vect{v}) = \frac{1}{V} \int \frac{d^3\vect{q}}{(2\pi)^3} \sum_{f} \bigl|M^{i,f}(\vect{q}, \vect{v}) \bigr|^2 \,2\pi\,\delta\bigl(E_f-E_i-\omega_{\vect{q}}\bigr) \,
\label{eq:rate}
\end{equation}
where $E_i$, $E_f$, and $\omega_{\vect{q}}$ are the initial, final, and transferred energies, respectively, $V$ is the volume of the detector, and $\vect{v}$ is the incoming DM velocity. 

The matrix element for the DM--phonon interaction reads
\begin{equation}\label{eq:matrix-element}
M^{i,f}(\vect{q}, \vect{v}) =   F_{\mathrm{DM}}(q)\sum_{l,j}\langle f |\, e^{i\vect{q}\cdot\vect{x}_{lj}}\,\widetilde{\cal V}_{lj} (-\vect{q},\vect{v})\, |i\rangle\,,
\end{equation}
assuming the material to be at the temperature of absolute zero, having no collective modes at the initial state $|i\rangle = |0\rangle$ and exciting to a single phononic band $\nu$ with momentum $\vect{k}$ leading to a final state $\langle f | = \langle \nu, \vect{k}|$.

Two distinct cases are considered when it comes to the comparison of the mass of the particle that mediates the DM--SM interaction with respect to the transferred momentum. These are encoded in the momentum-dependent DM form factor as:
\begin{align}
    F_{\mathrm{DM}}(q) &= \begin{cases}
    1  &\text{ for a heavy mediator,}\\
    \left(\frac{q_\mathrm{ref}}{q}\right)^2 &\text{ for a light mediator,}\\
    \end{cases}
\end{align}
where the reference momentum has been set to $q_\mathrm{ref} = \alpha m_e$ for the case of scattering with electrons and $q_\mathrm{ref} = m_\chi v_0$ for scattering with nucleons~\cite{Griffin:2019mvc}, where $v_0$ is the most probable DM velocity defined in the text that follows.

The Fourier transform of the DM--ion scattering potential for an ion $j$ within a primitive cell $l$ (and a position of $\vect{x}_{lj}$) can be expressed within the EFT formalism as
\begin{eqnarray}
\tilde{\cal V}_{lj}(-\vect{q},\vect{v}) &=& \sum_{\psi=\text{p,n,e}} 
c_1^{(\psi)} \langle N_\psi\rangle_{lj}\nonumber\\
&&\quad
+\,c_{5a}^{(\psi)} \,\frac{i\vect{q}}{m_\psi} \cdot\bigl(\vect{v}_\chi\times\vect{S}_\chi\bigr)\langle N_\psi\rangle_{lj}\nonumber\\
&&\quad
+\,c_{8a}^{(\psi)} \,\bigl(\vect{v}_\chi\cdot\vect{S}_\chi\bigr) \langle N_\psi\rangle_{lj}\nonumber\\
&&\quad
+\,c_{11}^{(\psi)} \frac{i\vect{q}}{m_\psi}\cdot\vect{S}_\chi\,\langle N_\psi\rangle_{lj}\nonumber\,,\nonumber\\
\label{eq:Vs}
\end{eqnarray}
since the contribution from the other effective operators is suppressed. The scalar couplings $c_i^{(\psi)}$ that regulate the strength of the interaction are included in the effective Lagrangian along with the interaction operators and DM (fermionic) non-relativistic states $\chi^\pm$ ($\psi^\pm$) as
\begin{equation}
\LL_\text{eff} \supset \sum_{i} \sum_{\psi=\text{p,n,e}} c_i^{(\psi)} \OO_i^{(\psi)} \chi^- \chi^+ \psi^- \psi^+\,,
\label{eq:lagrangian}
\end{equation}
and their non-zero combination determines the type of interaction a DM particle has with its SM counterpart.

The dark matter interaction rate per detector unit mass that is used to calculate detector sensitivities and exclusion limits is given by
\begin{equation}
R = \frac{1}{\rho_T}\frac{\rho_\chi}{m_\chi} \int d^3v\, f_\chi(\vect{v}) \,\Gamma(\vect{v})\,,
\label{eq:rate-total}
\end{equation}
where $\rho_\chi =0.4$\,GeV/cm$^3$~\cite{Trickle:2020oki} is the local DM density, $\rho_T$ is the density of the target material, and $f_\chi(\vect{v})$ is the DM velocity distribution, assumed to take the usual form of a truncated Maxwell-Boltzmann distribution
\begin{equation}
f_\chi(\vect{v}) = \frac{1}{N_0} e^{-(\vect{v}+\vve)^2/v_0^2} \,\Theta \bigl(\vesc-|\vect{v} +\vve|\bigr) \,,
\label{eq:maxwell-boltzmann}
\end{equation}
taking $v_0=230$\,km/s, $\vesc=600$\,km/s, $\ve=240$\,km/s~\cite{Trickle:2020oki} with $N_0$ set to unit-normalize the integral of the distribution.

The reference cross section throughout this work is defined as
\begin{equation}
\overline\sigma_\psi = \frac{(\mu_{\chi \psi} c_i^{(\psi)})^2}{\pi} \,,
\label{eq:cross-section}
\end{equation}
where $\psi$ denotes the struck fermion and $\mu_{\chi \psi}$ is the reduced mass of the fermion-DM system.

We focus on each of the allowed effective operators individually as well as on one DM model in particular---the dark photon. In this model, a new massive boson set by an $U(1)$ extension of the SM Lagrangian kinetically mixes with the ordinary photon. It has enjoyed great popularity in the DM direct detection community, serving as the first benchmark between various experiments, theoretical models, and experimental setups \cite{Fabbrichesi:2020wbt}. Following the approach of Ref.~\cite{Trickle:2020oki}, we model the dark-photon-like interaction by setting $c_1^{(e)} = -c_1^{(p)} = 1$.

\section{Magnetic moments arising from chiral phonons}\label{sec:dft-and-kappa}

We perform \textit{ab initio} calculations of phononic structures and chirality in several noncentrosymmetric MOFs (Fig. \ref{fig-crystal-structures}) using density functional theory (DFT) and density functional perturbation theory (DFPT) (see Sec.~\ref{sec:methods} for details). Three MOFs with relatively small numbers of atoms in the unit cell were studied using a $2 \times 2 \times 2$ grid in $\mathbf{q}$ (phonon wavevector) space. These three MOFs have the chemical formulae CuCl(pyrimidine) (CuCl(pyr), space group $Pma2$)~\cite{natherSynthesis2004}, AgCl(1,10-phenanthroline) (AgCl(phen), space group $C_2$)~\cite{odokoCatena2004}, and Sr(\textsc{l}-tartrate) (Sr(tar), space group $C222_1$)~\cite{appelhansPhase2009}.

The results of these calculations (\textit{vide infra}) demonstrate that the high degree of real-space structural complexity of MOFs leads to low band dispersion, making the impact of the $\mathbf{q}$-space resolution on the projected detector efficiency negligible. We are therefore able to extend our study by performing calculations at the $\Gamma$ ($\mathbf{q} = (0 ~0 ~0)$) point of three, more complex, MOFs: Li$_2$(\textsc{l}-tartrate) (Li$_2$(tar), space group $P2_1$)~\cite{yeungChiral2013}, Zn(NH$_4$(formate)$_3$ (Zn(NH$_4$)(fmt)$_3$, space group $P6_3$)~\cite{xuDisorderOrder2010}, and Cd(guanadinum)(formate)$_3$ (Cd(Gua)(fmt)$_3$, space group $Cc$)~\cite{collingsCompositional2016}. 

In order to compare MOFs with conventional inorganic materials, we include  in our analysis GaAs (space group $F\bar{4}3m$) and LiNbO$_3$ (space group $R3c$), whose DFT-calculated dynamical matrices we obtain from the literature~\cite{jainCommentary2013, osti_1200591, uedaChiral2025}. These materials are of particular interest because optical measurements observe phonon magnetism in LiNbO$_3$~\cite{biggsUltrafast2025}, and suggest it in $n$-doped GaAs~\cite{wysmolekCoupled2006}. We also include the MOF studied in our previous work~\cite{Romao:2023zqf}: InF$_3$($4,4'$-bipyridine) (InF$_3$(bpy), space group $I222$) \cite{petrosyants2010organometallic}.

The phononic structures allow us to assess the potential of these materials as DM detector candidates based on the energies of the modes (which determine the possible DM-phonon interactions, following Eqs.~(\ref{eq:rate})-(\ref{eq:rate-total})), and their chirality, which causes the phonons to carry angular momentum. The angular momenta ($\mathbf{J}$) of the phonons are obtained from their eigenvectors ($\varepsilon$) using the circular polarization operator ($\hat{S}$)~\cite{zhang2015chiral}:
\begin{align}
    \label{eq-pam}
     \mathbf{J}_{n,\mathbf{q}} = \sum_{\alpha,n,\mathbf{q}} \mathbf{J}_{\alpha,n,\mathbf{q}} = \hbar \sum_{\alpha,n,\mathbf{q}} \bra{\epsilon_{\alpha,n,\mathbf{q}}} \hat{S} \ket{\epsilon_{\alpha,n,\mathbf{q}}},
\end{align}
where $\alpha$ and $n$ index the atoms in the unit cell and the phonon bands, respectively. Phonon band structures of GaAs, LiNbO$_3$, CuCl(pyr), and AgCl(phen), with the bands coloured by the magnitude of $\mathbf{J}$, are shown in Fig.~\ref{fig-band-structures}. The Cartesian components of the vector $\mathbf{J}$ are plotted in the Appendix (Sec.~\ref{sec:appendix}). 

The phonon band structures of the MOFs CuCl(pyr), AgCl(phen), and Sr(tar) contain modes with negative energy, indicating the presence of phonon instabilities at $0~\mathrm{K}$. These unstable modes are excluded from our calculations of the detector sensitivity. Such instabilities could indicate a displacive phase transition on cooling, or the presence of crystallographic disorder, which could be random or correlated~\cite{simonovDesigning2020}. Any displacive phase transition on cooling would not add symmetry elements to the space group, and therefore could not remove the chirality of the phonons. Disorder, which is very common in MOFs~\cite{meekelCorrelated2021}, would maintain the average structure of the crystal, and could have beneficial effects by enhancing the lifetimes~\cite{suzukiDisorder2025} and localization~\cite{beardoResonant2024} of chiral phonons. Unfortunately, due to the large number of potential multiplications of the unit cell corresponding to the different instabilities, it was not plausible to explore this topic further using DFT.

\begin{figure*}[ht]
    \centering
    \includegraphics[width=0.99\textwidth,trim=1cm 1cm 1cm 0cm,clip]{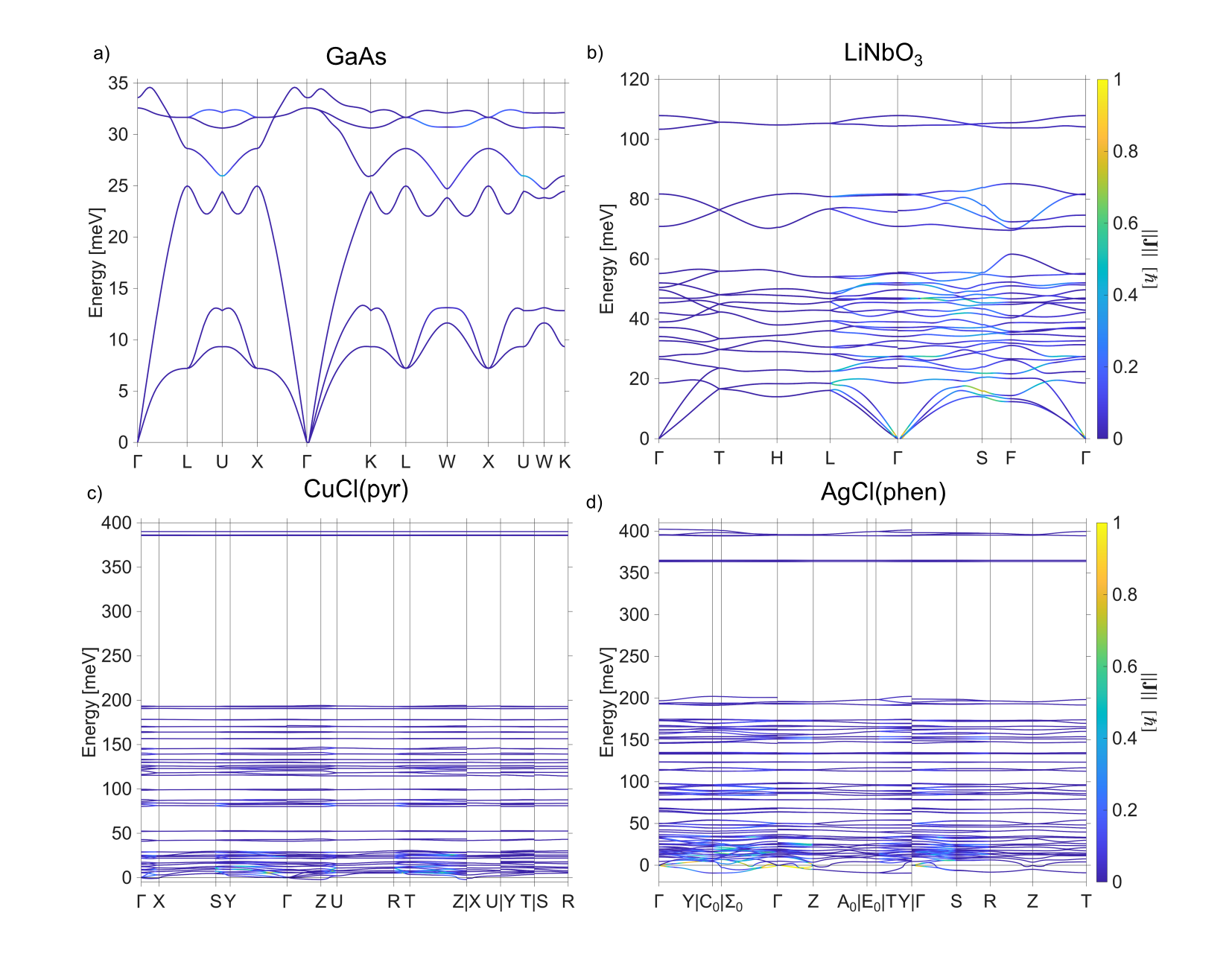}
  \caption{Calculated phonon band structures of  GaAs (panel \textbf{a}), LiNbO$_3$ (panel \textbf{b}), CuCl(pyr) (panel \textbf{c}), and AgCl(phen) (panel \textbf{d}), with bands coloured according to their phonon chirality, as quantified by the magnitude of the phonon angular momentum ($||\mathbf{J}||$). Special points in and paths through the Brillouin zone were chosen following the literature \cite{hinuma2017band}. The dynamical matrices of GaAs and LiNbO$_3$ were obtained from the literature~\cite{osti_1200591, uedaChiral2025}.}
  \label{fig-band-structures}
\end{figure*}

Beyond their impact on the DM--phonon interaction cross-section (see Sections~\ref{sec:dm-theory} and~\ref{sec:sensitivity}), several features of the phonon band structures affect the suitability of the materials as chiral phonon-based DM sensors~\cite{Romao:2023zqf}. Such a sensor requires long-lived phonons, localized near a magnetometer. The magnetic flux sensitivity of SQUID magnetometers is generally limited by flux noise, and so this sensitivity scales with $\sqrt{t}$ \cite{ketchenDesign1989, stormUltrasensitive2020}. Phonons with longer lifetimes therefore require smaller magnetic moments to be detectable. Long (\textmu s--s) phonon lifetimes are only possible in the lowest-energy phonon band (the slow transverse acoustic band), as phonons in higher-energy bands decay into this band on a ps--ns timescale ~\cite{maccabe2020nano, maris1993anharmonic}. The stability of phonons in the lowest-energy band is due to the selection rules for their decay, which require conservation of energy and momentum~\cite{maris1993anharmonic}; therefore flatter acoustic bands increase the volume of reciprocal space with stable phonons. Phonon localization is also improved by reducing the phonon group velocity, which again implies that flatter bands are desirable. The MOFs have significantly flatter phonon bands than the inorganic materials, as expected due to their porosity~\cite{kamencek2019understanding}.

The most important quality of chiral phonons for sensing applications is their magnetic nature. Unfortunately, this is also the most difficult to predict, as there is not yet a theoretical consensus regarding the origin of giant phonon magnetic moments in diamagnetic materials~\cite{juraschekChiral2025, shabalaAxial2025}. However, a common factor between the proposed mechanisms is their dependence on the angular momentum of the atoms participating in the phonon. A somewhat general expression for the magnetic moment ($\mathbf{M}$) of a phonon can be given as: 
\begin{align} 
    \label{eq-magmo}
    \mathbf{M}_{n,\mathbf{q}} = \sum_{\alpha} \mathbf{J}_{\alpha,n,\mathbf{q}}  \frac{\mathbf{Z}_\alpha^\star}{2m_\alpha}
\end{align}
where $\mathbf{Z}_\alpha^\star$ is an effective charge, containing contributions from, for example, motions of the Born effective charges~\cite{juraschek2017dynamical, juraschek2019orbital}, topological currents~\cite{ren2021phonon, hernandezObservation2023}, and dynamical quadrupoles \cite{zabaloRotational2022}. The magnetic moments arising from the contributions of the Born effective charges for materials included in the present study are shown in the Appendix (Sec.~\ref{sec:appendix}). 

\begin{figure*}[ht]
    \centering
    \includegraphics[width=0.99\textwidth,trim=0cm 0cm 0cm 0cm,clip]{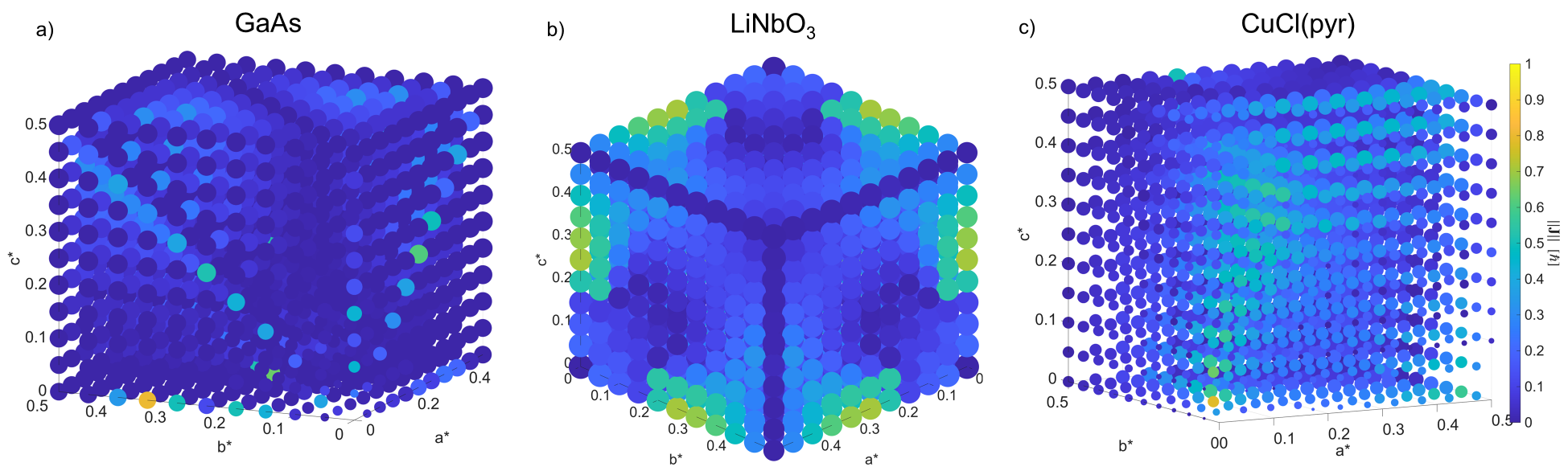}
  \caption{Phonon angular momentum ($\mathbf{J}$) in the lowest-energy acoustic band  of GaAs (panel \textbf{a}), LiNbO$_3$ (panel \textbf{b}), and CuCl(pyr) (panel \textbf{c}). Phonon modes in this band are displayed as points on a $10 \times 10 \times 10$ grid in reciprocal space, with the colour of the point showing the magnitude of $\mathbf{J}$ and the size of the point displayed in proportion to the phonon energy. The expectation value of the phonon angular momentum in this band ($\langle ||\mathbf{J}_1|| \rangle$) is used to estimate $\kappa$, the probability of the decay of an excitation into at least one detectable chiral phonon. For GaAs, $\kappa = 0.07$, for LiNbO$_3$  $\kappa = 0.22$, for CuCl(pyr) $\kappa = 0.16$. Unstable modes in CuCl(pyr) are not shown.}
  \label{fig-cubes}
\end{figure*}

Therefore, as phonon angular momentum is a prerequisite for phonon magnetism, we focus on the presence and distribution of this quantity in the stable acoustic band. In Fig.~\ref{fig-cubes}, we plot the phonon modes in this band on a $10 \times 10 \times 10$ grid of $\mathbf{q}$ points in reciprocal space. We use such a grid to visualize the phononic properties at arbitrary points because important chiral features can be present at points away from the high-symmetry lines shown in the phonon band structures. The regions of reciprocal space containing chiral phonons in the lowest energy band are marked in Fig.~\ref{fig-cubes} by green-yellow points.

The variation in phonon chirality seen between the three materials can be understood by reference to their crystal structures. The atoms in GaAs are packed tightly, and therefore circular motion of an atom creates forces on its neighbours, increasing the energy of the chiral phonons out of the lowest band. By contrast, CuCl(pyr) shows phonon chirality in a much broader region of reciprocal space, which can be attributed to its framework structure, which contains empty space in the unit cell and topologically underconstrained atoms. Interestingly, LiNbO$_3$ also shows a significant amount of phonon chirality in its lowest energy band, possibly because the phonons in this band have higher energies than their counterparts in GaAs and CuCl(pyr).

Our simulations of the DM detector sensitivity require an estimate of the probability that an excitation produces a detectable phonon ($\kappa$). Given that all excitations will result in the production of at least one phonon in the lowest energy band, we consider the probability that this phonon is detectable, following the approach of our earlier work~\cite{Romao:2023zqf}. We conservatively assume that each excitation produces only one phonon, which is randomly located within the lowest energy band. We also assume the probability of detection of such a phonon to be proportional to its phonon angular momentum, with a phonon of maximum angular momentum ($\hbar$) having a detection probability of 1 (\textit{i.e.,} we assume the magnetometers are sufficiently sensitive to detect such phonons). 

These assumptions allow us to estimate $\kappa$ as $\kappa = \langle ||\mathbf{J}_1|| \rangle/\hbar$, the expected chirality of the lowest energy phonon band. We find $\kappa = 0.07$ in GaAs, $\kappa = 0.22$ in LiNbO$_3$, and $\kappa = 0.16$ in CuCl(pyr). The value  of $\kappa$ in CuCl(pyr) is very similar to that found in InF$_3$(bpy) ($\kappa = 0.15$)~\cite{Romao:2023zqf}, and so we assume that the value is similar for the other MOFs (where calculations were performed only at $\Gamma$ or limited by the presence of phonon instabilities), and adopt $\kappa = 0.16$ for all MOFs in this study.

Fortunately, $\kappa$ appears as a linear scaling parameter in the logarithmic phase space of detector sensitivity, and so the assumptions we make regarding $\kappa$ do not affect our conclusion that the chiral phonon sensing paradigm offers significant potential advantages over other current and proposed DM detector designs~\cite{Romao:2023zqf}. Our modeling assumes that we can identify a material with large acoustic phonon magnetic moments at arbitrary $\mathbf{q}$, whereas experimental reports of phonon magnetism to date have primarily focused on optic phonons at $\Gamma$, as these are the easiest to manipulate using light, and have potential technological applications in the emerging field of quantum printing~\cite{aeppliQuantum2025}. Therefore, in conjunction with continuing theoretical efforts to predict phonon magnetism \textit{ab initio}, further methodological development in the sensing of phonon magnetism would help us develop more accurate models of the detector sensitivity, and we propose two such methods in Sec.~\ref{sec:measurement}.

\section{Direct measurement of the magnetic moments}\label{sec:measurement}
In our previous work, we proposed a detection scheme using a material that hosts magnetic chiral phonons equipped with sensitive magnetometers at the sides~\cite{Romao:2023zqf}. After an interaction of the material with a DM particle (or any other type of ionizing radiation), low-energy stable acoustic phonons of both chiralities are created in equal numbers proportional to the energy transfer. 

The thermal phonon Hall effect, recently shown to be universal in all crystals~\cite{jinDiscovery2025}, will then be used to lead phonons of opposite handedness to the opposite sides of the sensor. The phonons will then localize at the surface of the material, adding their individual magnetic moments to the total field that the magnetometers will be exposed to. If the interaction deposited enough energy to trigger the magnetometers, an event will be recorded.

Before we can proceed to the fabrication of such a prototype, we must identify the ideal material that hosts stable magnetic chiral phonons generating fields detectable with a magnetometer. Although the magnetic properties of phonons have been observed indirectly through the Faraday effect~\cite{hernandezObservation2023}, Kerr rotation~\cite{basiniTerahertz2024}, and the Zeeman splitting of the phononic states~\cite{BaydinPbTe}, a direct readout with a mounted magnetometer has not yet been reported.

\begin{figure*}[ht]
    \centering
    \includegraphics[width=0.49\textwidth]{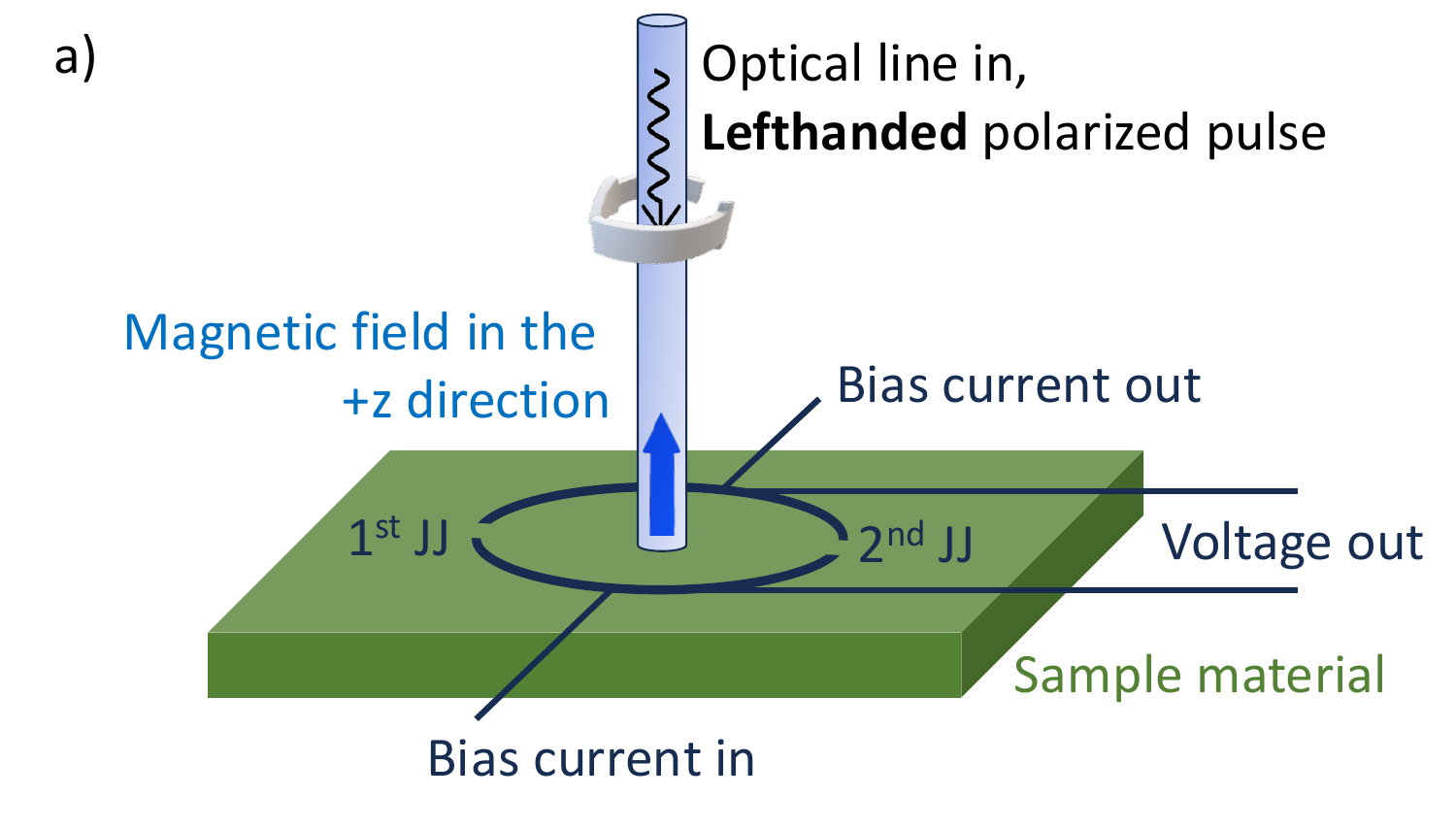}
    \includegraphics[width=0.49\textwidth]{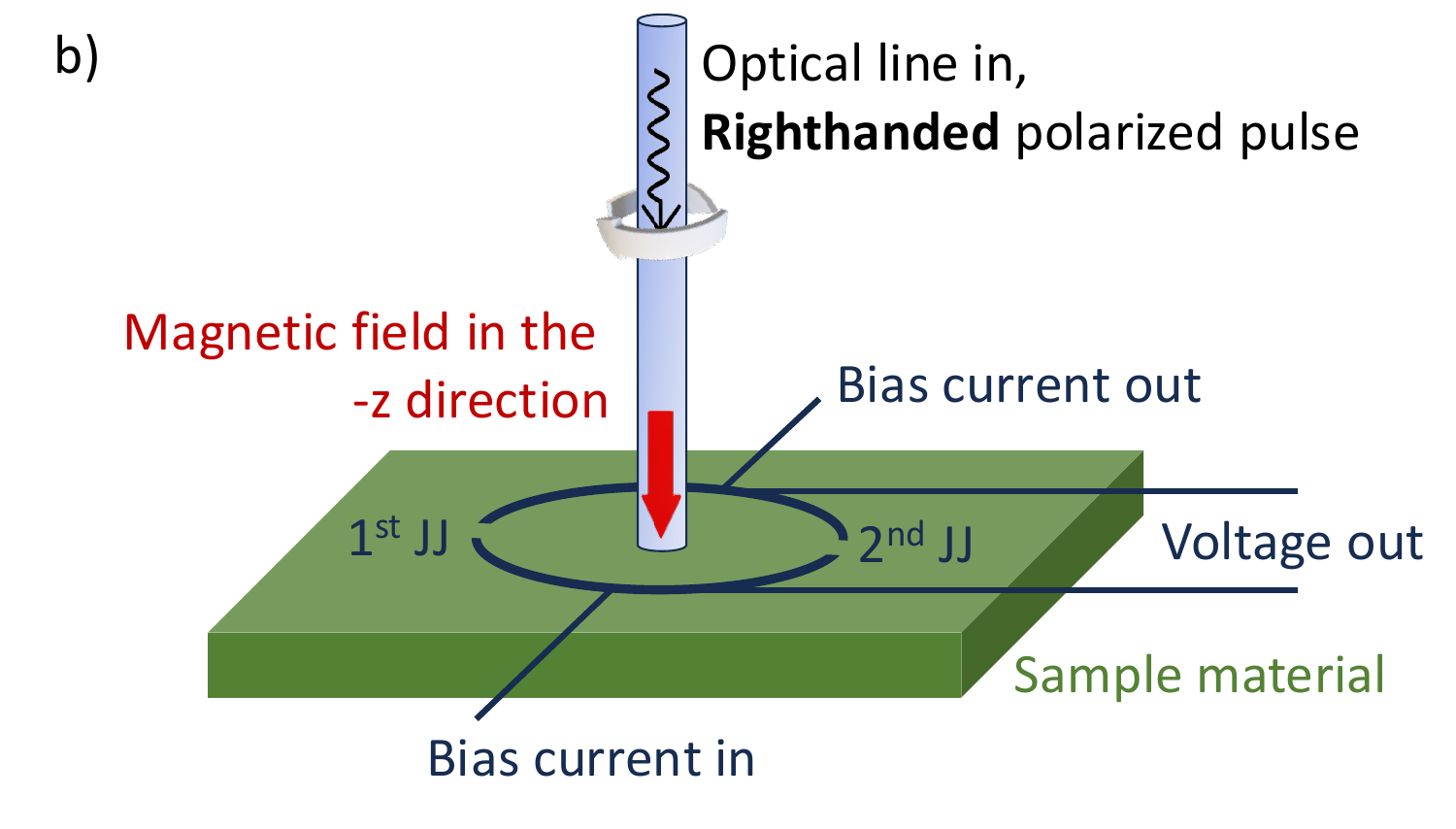}
    \includegraphics[width=0.49\textwidth]{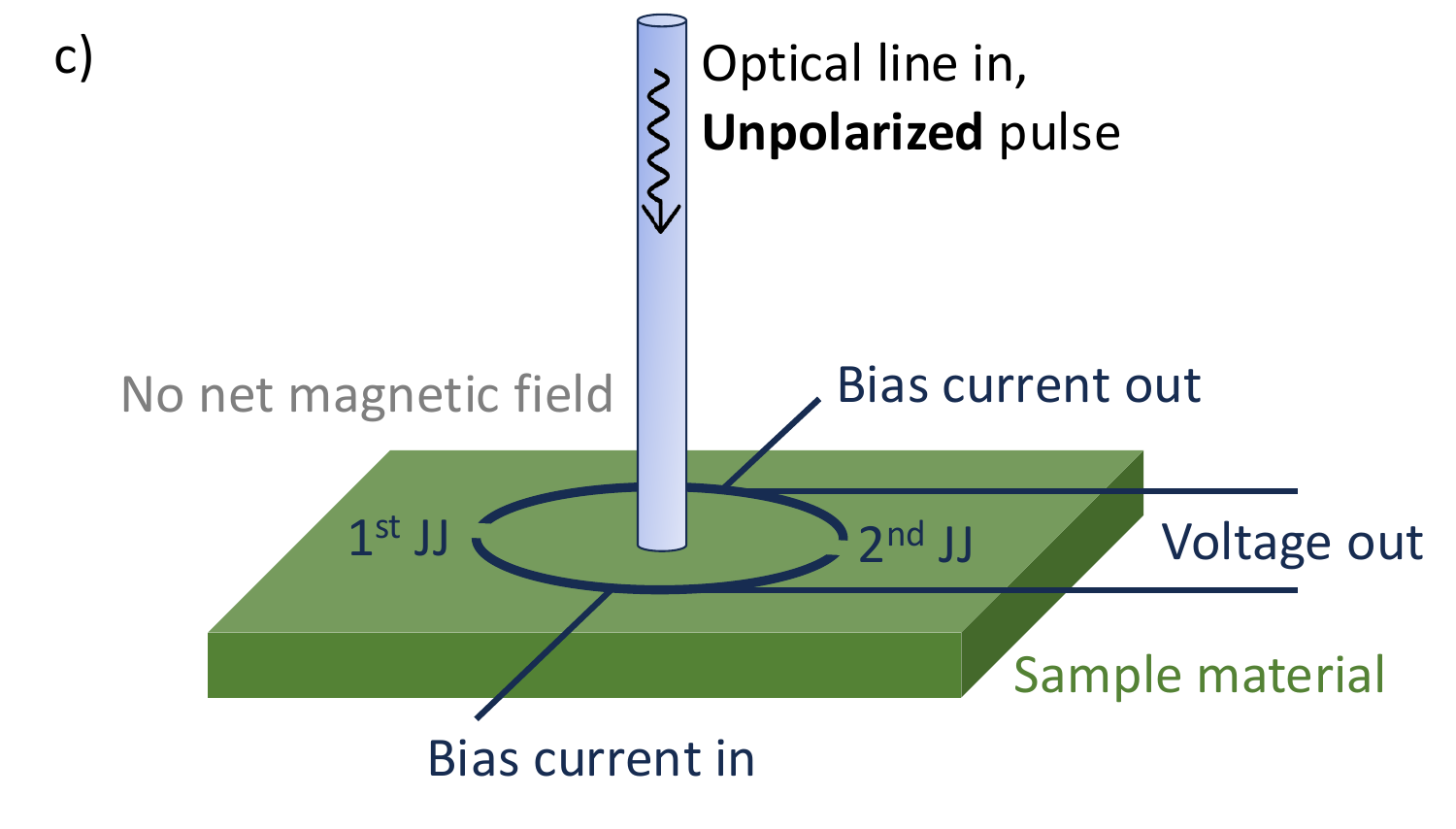}
    \includegraphics[width=0.49\textwidth]{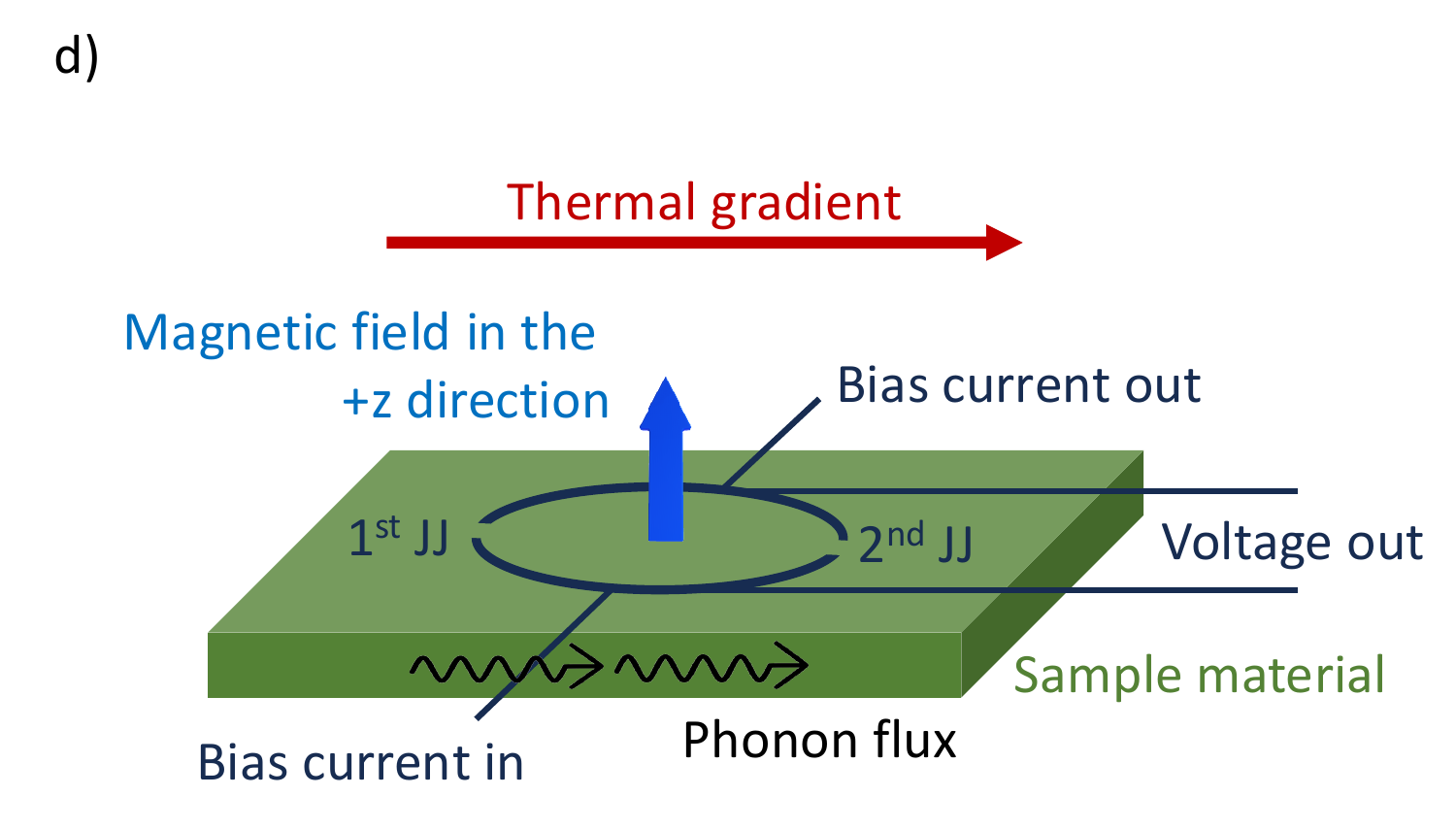}
  \caption{Testing setup for a chiral phonon sensor prototype. A material which hosts magnetic chiral phonons is mounted with a surface magnetometer (in our example, an assymmetric SQUID in a 4-probe setup). Chiral phonons are excited by a circularly polarized laser pulse and the magnetic field fluctuation they generate is read out with the magnetometer (panel \textbf{a}). Upon switching polarization of the laser, the magnetic field induced by chiral phonons is inverted (panel \textbf{b}). An unpolarized laser pulse induces identical populations of phonons of both chiralities, resulting in a suppressed magnetic response of the sample for benchmarking purposes (panel \textbf{c}). A population of chiral phonons with a preferred handedness can also be excited by application of a thermal gradient driving a preferential direction of phononic transport (panel \textbf{d}).}
  \label{fig-testing-prototype}
\end{figure*}

To assess the feasibility of our approach, we propose to fabricate a magnetic fluctuation detector based on an asymmetric SQUID loop on top of the studied material (Fig.~\ref{fig-testing-prototype}). This is a relatively simple nanofabrication process, which relies on ex-situ deposition of a superconducting thin film (e.g. NbTiN on LiNbO$_3$~\cite{Telkamp:2025jkd}) and defining a loop with two asymmetrical Josephson junctions by electron beam lithography and dry/wet etching. Such a device is compatible with 4-probe measurement of current–phase relations (CPR)~\cite{doi:10.1021/acs.nanolett.3c01970}. The advantage of CPR is that it is not sensitive only to a global flux, but also to phase perturbations within the SQUID loop. That allows detection of extremely small magnetic-field fluctuations—down to the level of individual Bohr-magneton–scale moments~\cite{vasyukov2013scanning}. 

Naturally, the operational temperature range is limited by the critical temperature of the selected superconductor. However, it is in any case desirable to operate at mK temperatures due to suppression of thermal noise. Once the material is mounted with a magnetometer and cooled down, we need to selectively excite a population of phonons of a given handedness in order to generate magnetic fluctuations within the SQUID loop. We identify two options for selectively exciting a specific handedness:

\textit{i}) We can expose the sample to a circularly polarized laser pulse through an optical line or window into the cryostat (Fig.~\ref{fig-testing-prototype}~a). The conservation of angular momentum implies that phonons of the same handedness as the laser pulse will be preferentially excited \cite{quartz}. Flipping the chirality of the pulse will reverse the population of the phonons, inverting the field picked up by the magnetometer (Fig.~\ref{fig-testing-prototype}~b). If the sample is then exposed to an unpolarized laser pulse, phonons of both chiralities will be created in identical populations, averaging out the magnetic response, leading to no observable signal (Fig.~\ref{fig-testing-prototype}~c). This would allow us to establish a baseline and attribute signal from circularly polarized excitations to the chiral phonons.

The laser itself does not need to be in resonance with a specific optic phonon, as non-resonant interactions will generate heat in the form of acoustic phonons. While resonant excitation would be significantly more efficient, passing circularly polarized THz radiation into the cryostat could be challenging. We note that, in polar materials (such as LiNbO$_3$ and CuCl(pyr)), non-resonant interactions can create ferrons \cite{choeObservation2025}, which are quasiparticle excitations of the bulk polarization coupled to optic phonons \cite{bauerPolarization2023}. Circularly polarized ferrons, coupled to circularly polarized phonons, are predicted to also create magnetic effects \cite{polsMultiferrons2025, tangMultiferroiclike2025}, and therefore are another potential source of signal.

\textit{ii}) Applying a thermal gradient perpendicular to the device orientation will cause a preferential flow of phonons of one handendess along the thermal gradient (Fig.~\ref{fig-testing-prototype} d) \cite{funatoChiralityinduced2024}. Heat is predominantly carried by phonons in the lowest-energy acoustic band, which in noncentrosymmetric materials can host chiral phonons \cite{Romao:2023zqf}. In materials with time-reversal symmetry, reversing the direction of propagation of the phonons reverses their chirality. Reversing the thermal gradient reverses the flow of phonons, thereby resulting in an opposite magnetization of the surface of the sample. In principle, the presence of a thermal gradient could interfere with the operation of the SQUID by creating a differential critical current between the two arms. This would limit the magnitude of the thermal gradient which could be used, or it could necessitate the use of an alternate design of the magnetometric readout, for example by employing a SQUID not in thermal contact with the sample, or by using an NV magnetometer.

The use of a thermal gradient to create a magnetization requires the phonons to possess magnetic moments that are not collinear with their direction of propagation. Only achiral noncentrosymmetric materials (\textit{e.g.} GaAs, LiNb)$_3$, CuCl(pyr), Cd(Gua)(fmt)$_3$) are permitted by symmetry to have phonons with components of the angular momentum orthogonal to their propagation~\cite{yangCatalogue2025}. The phonon magnetic moment is, in simple cases, expected to be aligned with the phonon angular momentum, however, in the general case, the relative orientation of the phonon propagation and magnetic moment requires a mechanistic understanding of the phonon magnetism ~\cite{chaudharyAnomalous2025}.

We propose these methods of measurement in order to demonstrate the feasibility of identifying materials with the properties required for our proposed sensor. We can use this approach to verify the presence of observable magnetic fields induced by chiral phonons in our test material, benchmark its sensitivity, and then proceed to the fabrication of a larger device readout using the phonon Hall effect for separation and localization of opposite handednesses of phonons. The expected sensitivity of such a sensor used for the direct detection of dark matter will be discussed in the next section.

\section{Detector sensitivity}\label{sec:sensitivity}

To study the expected sensitivity of the proposed chiral phonon detector, we study both a commonly explored interaction model---the dark photon model---as well as the all-operator approach, considering more exotic interaction types. The exposure we considered throughout this work is kg$\cdot$yr with 95\% confidence limits assuming zero background (or equivalently a minimum of three recorded events). The minimal energy threshold of a phononic excitation required to record an event was taken as 1\,meV, essentially allowing all excitations away from the $\Gamma$-point of a sufficient chirality to be read out \cite{maris1993anharmonic}.

\begin{figure*}[ht]
    \centering
    \includegraphics[width=0.48\textwidth]{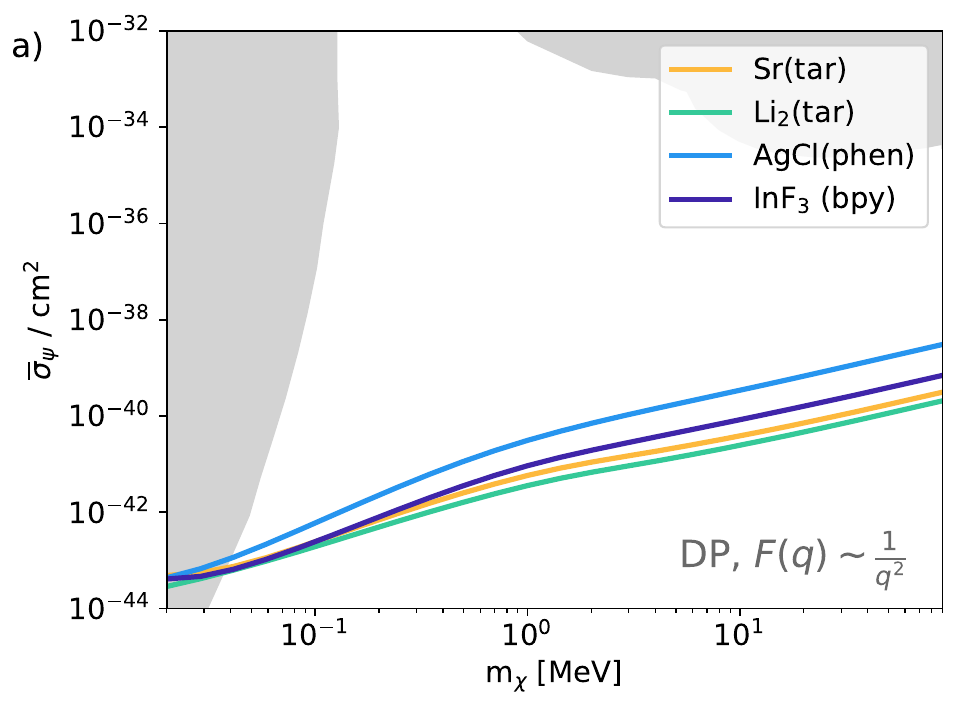}
    \includegraphics[width=0.48\textwidth]{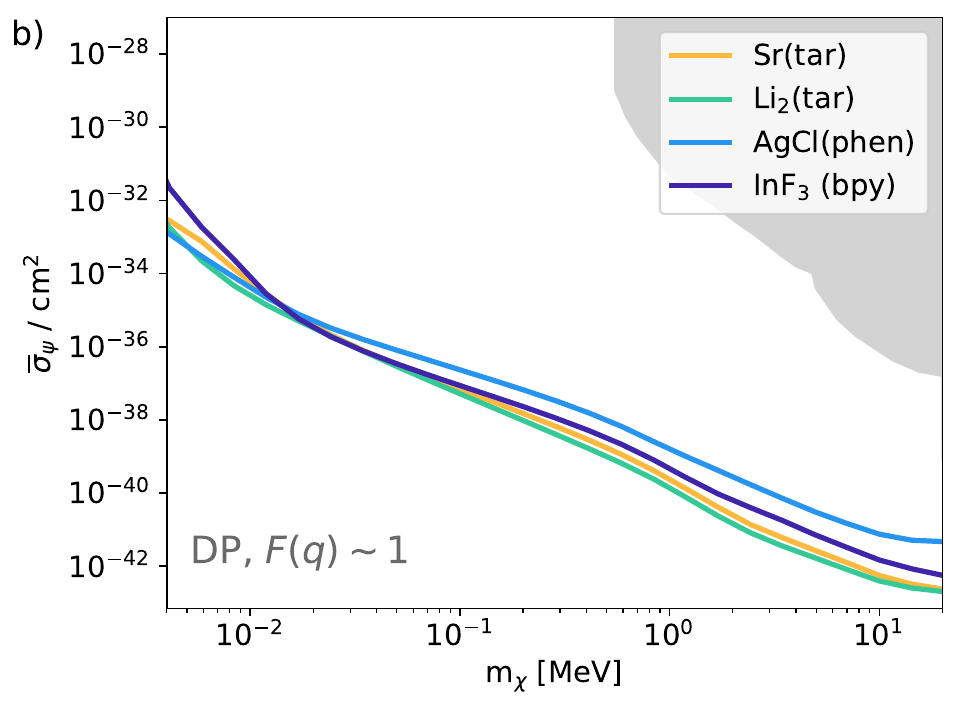}
    \includegraphics[width=0.48\textwidth]{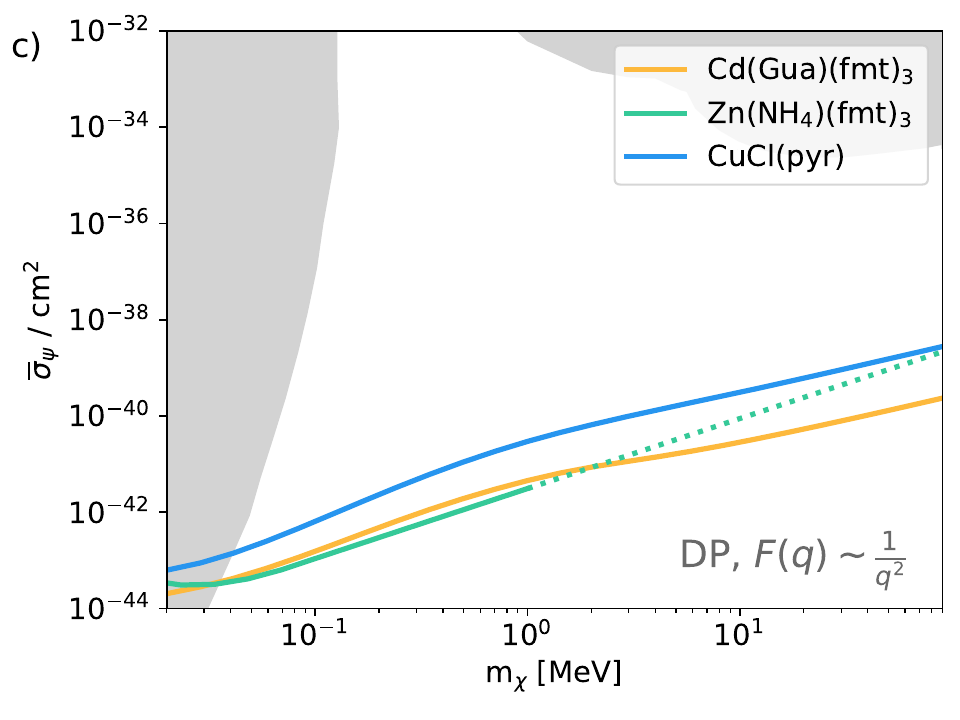}
    \includegraphics[width=0.48\textwidth]{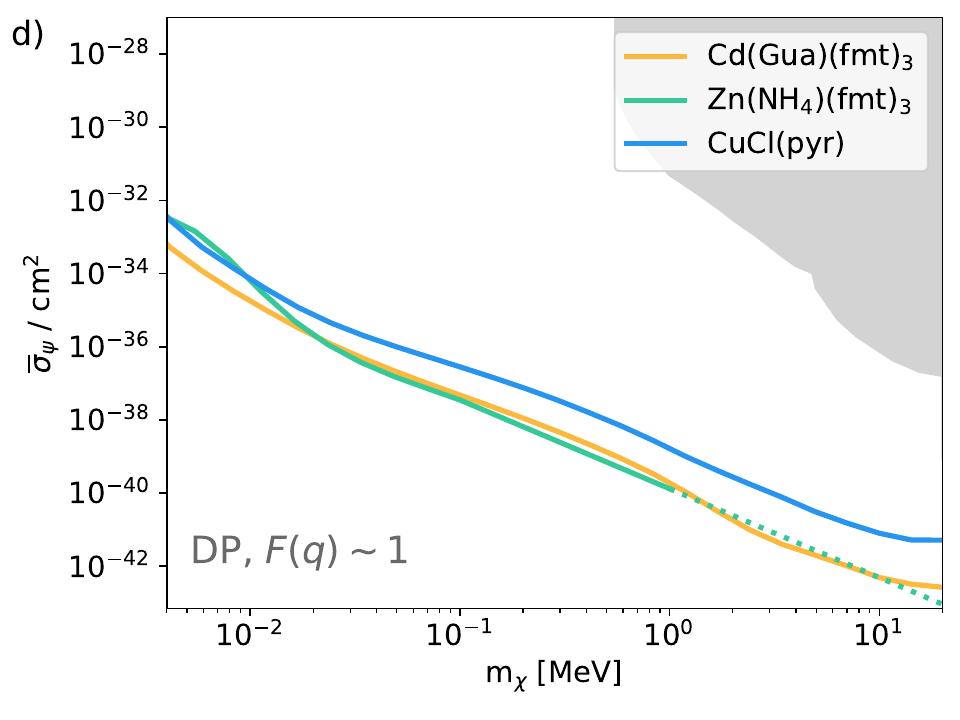}

  \caption{The dark photon interaction sensitivity for various MOF targets with kg$\cdot$yr exposure and 95\% confidence limits assuming zero background events, shown in terms of the interaction cross-section ($\bar\sigma_\psi$) and DM particle mass ($m_\chi$). Interactions with a light mediator are shown in the left column (panels \textbf{a} and \textbf{c}), while those with a heavy mediator are shown in the right column (panels \textbf{b} and \textbf{d}) (see Sec.~\ref{sec:dm-theory} for details). The top row (panels \textbf{a} and \textbf{b}) shows the first four selected MOFs, while the bottom one (panels \textbf{c} and \textbf{d}) shows the remaining three. The gray areas are those excluded by previous direct detection experiments~\cite{Essig:2015cda, SENSEI:2019ibb, DarkSide:2018ppu, DAMIC:2019dcn} and by constraints from astrophysical observations~\cite{Chang:2019xva}. The proportion of the chiral phase space has been set for all MOFs to $\kappa=0.16$. Data for InF$_3$(bpy) are reproduced from the literature~\cite{Romao:2023zqf}.}
  \label{fig-dark-photon-sensitivity-DP-MOFs}
\end{figure*}

\begin{figure*}[ht]
    \centering
    \includegraphics[width=0.48\textwidth]{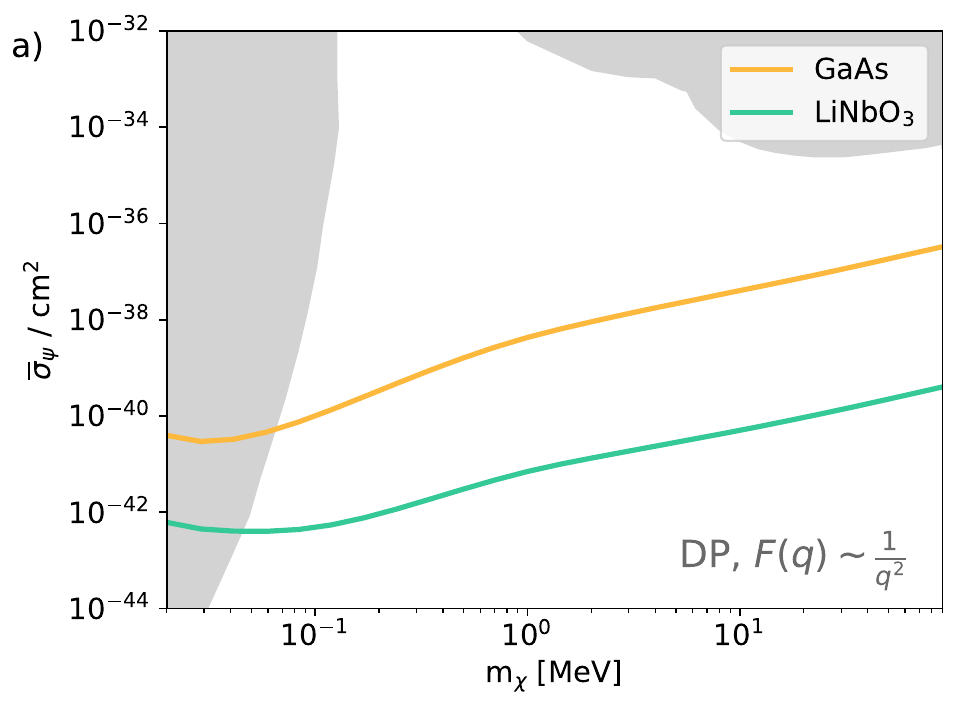}
    \includegraphics[width=0.48\textwidth]{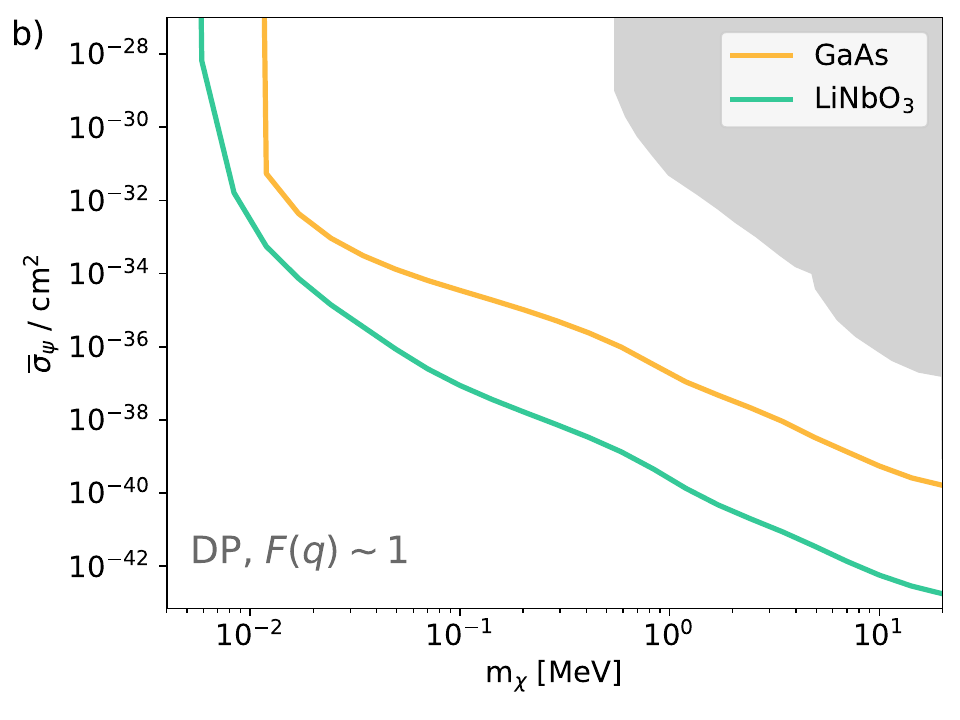}
  \caption{Same as Figure~\ref{fig-dark-photon-sensitivity-DP-MOFs}, but for crystalline GaAs and LiNbO$_3$ with the light mediator interaction shown in panel \textbf{a} and heavy mediator in panel \textbf{b}. The proportion of the chiral phase space has been set as $\kappa=0.07$ for GaAs and $\kappa=0.22$ for LiNbO$_3$.}
  \label{fig-dark-photon-sensitivity-DP-crystals}
\end{figure*}

\begin{figure*}[ht]
    \centering
    \includegraphics[width=0.48\textwidth]{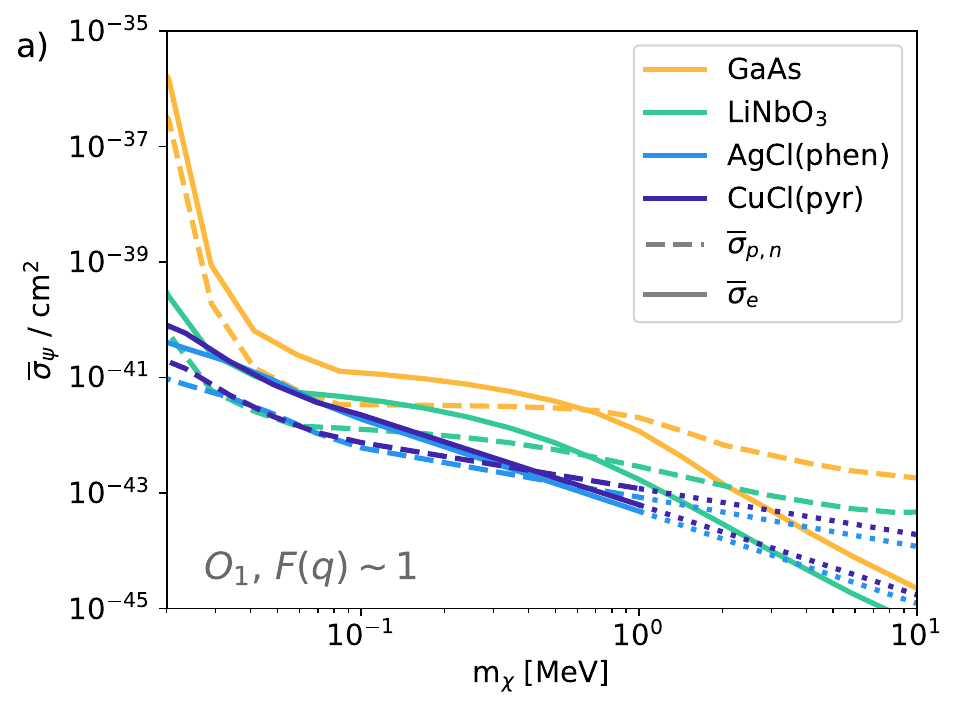}
    \includegraphics[width=0.48\textwidth]{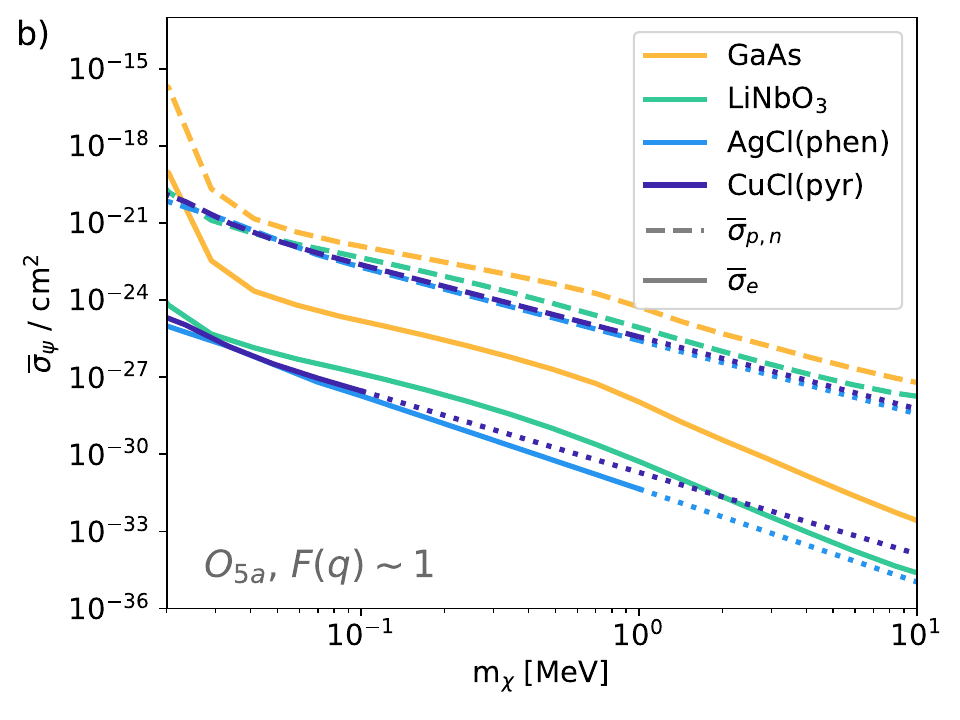}
    \includegraphics[width=0.48\textwidth]{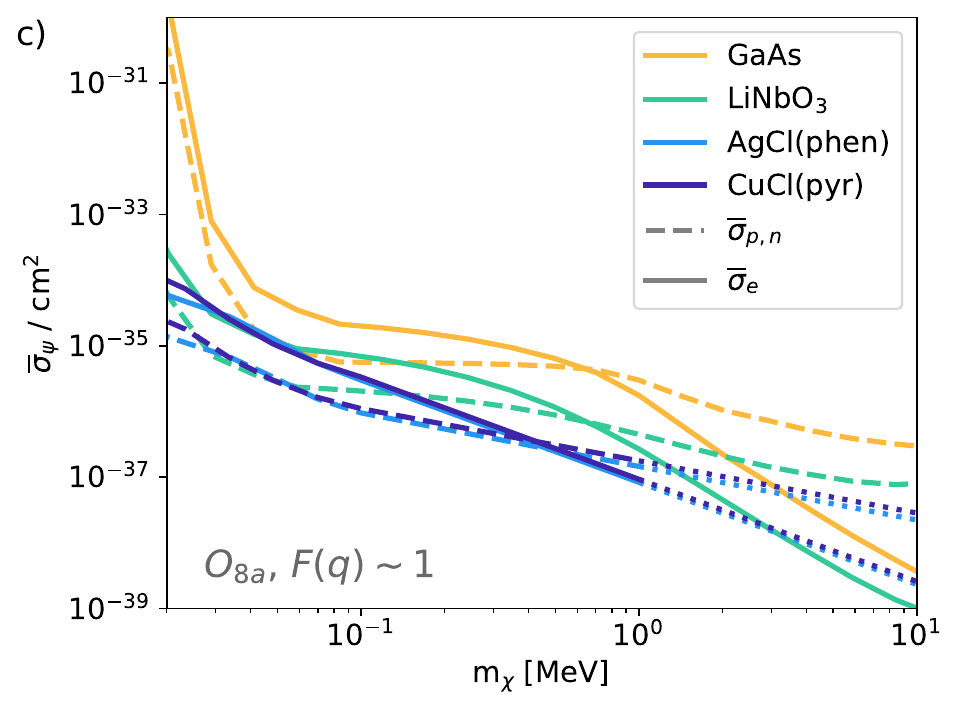}
    \includegraphics[width=0.48\textwidth]{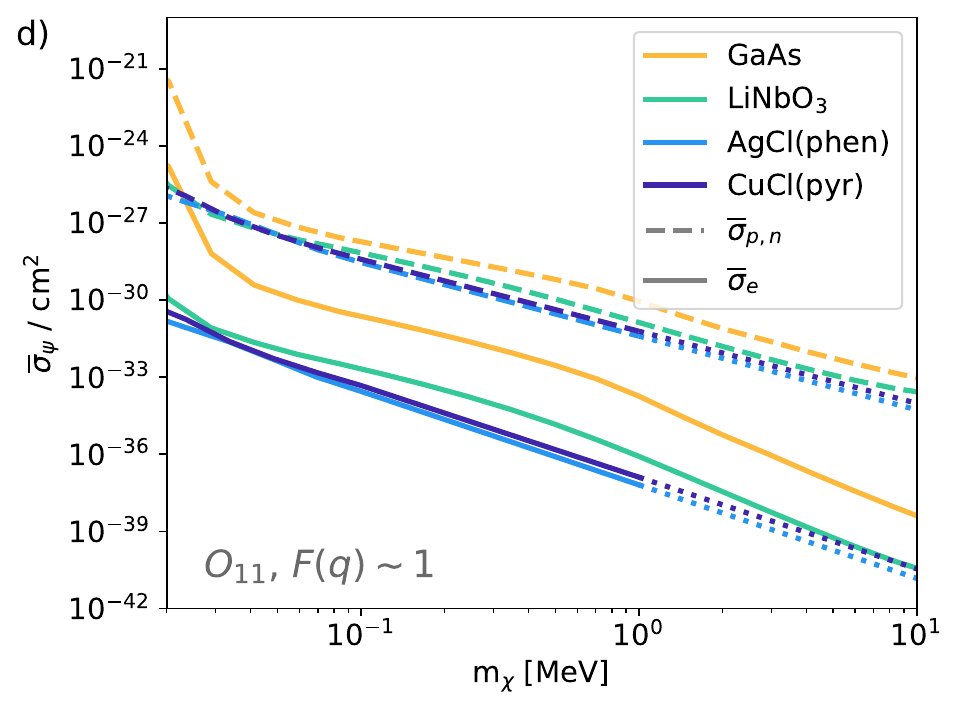}

  \caption{The leptophilic (solid) and hadrophilic (dashed) interaction sensitivity of selected MOFs and inorganic crystals for operators $\mathcal{O}_1$, $\mathcal{O}_5$, $\mathcal{O}_8$, and $\mathcal{O}_{11}$, as defined in Eq.~(\ref{eq:operators}). As the interaction phase space grows quickly with the increasing mass of the DM particle $m_\chi$ (which increases the number of kinematically allowed transitions), the computational requirements can become unfeasible. When this occurs, we outline a power-law-extrapolated rate with a dotted line. The proportion of the chiral phase space has been set as $\kappa=0.07$ for GaAs, $\kappa=0.22$ for LiNbO$_3$, and $\kappa=0.16$ for the two MOFs.}
  \label{fig-operators-sensitivity}
\end{figure*}

\begin{figure}[ht]
    \centering
    \includegraphics[width=0.48\textwidth]{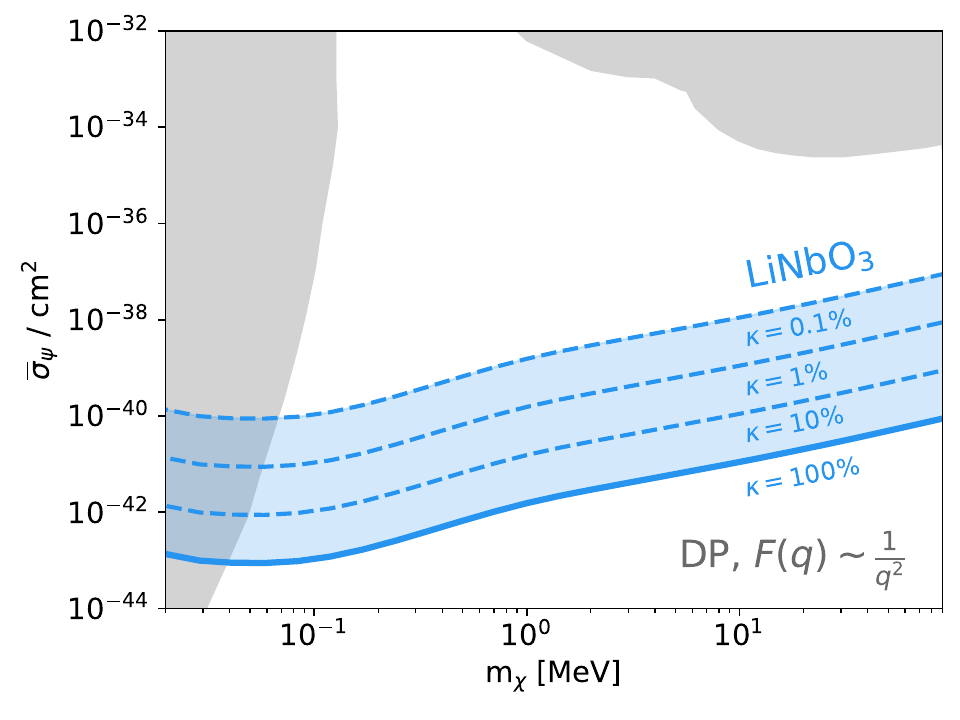}
  \caption{The influence of the proportion of the chiral phase space $\kappa$ on the projected sensitivity of a chiral phonon dark matter detector made of LiNbO$_3$. The dark photon model with a light mediator was taken as an example interaction type. Exposure of  kg$\cdot$yr was assumed with 95\% confidence limits and zero background. We can see that the proportion of the available magnetic chiral phonon phase space $\kappa$ linearly affects the resulting detector sensitivity. The gray shaded region has been excluded by past experiments and astrophysical observations~\cite{Essig:2015cda, SENSEI:2019ibb, DarkSide:2018ppu, DAMIC:2019dcn, Chang:2019xva}.}
  \label{fig-dark-photon-phasespace}
\end{figure}

Figs.~\ref{fig-dark-photon-sensitivity-DP-MOFs}--\ref{fig-dark-photon-phasespace} show the reach of chiral phonon-based detectors (\textit{i.e.}, the ability of the detector to rule out certain potential properties of DM) in terms of their sensitivity, as expressed by the interaction cross-section ($\bar\sigma_\psi$, see Eq.~(\ref{eq:cross-section})) and the DM particle mass ($m_\chi$). Fig.~\ref{fig-dark-photon-sensitivity-DP-MOFs} shows the expected reach for the studied MOFs in the dark photon approach, both in the limits of a light (left column) and heavy (right column) mediator, while Fig.~\ref{fig-dark-photon-sensitivity-DP-crystals} shows the same for a crystalline GaAs and LiNbO$_3$.  The sensitivity to either one of the effective operators expected to be non-zero within our framework is shown in Fig.~\ref{fig-operators-sensitivity} in the heavy mediator limit for a subset of the studied materials, comparing MOF targets to the inorganic crystals.

The phase space of all allowed transitions caused by an incident DM particle grows rapidly with the particle mass $m_\chi$. Since the maximum transferred momentum $q$ grows linearly with increasing DM mass, the integral over the allowed phase space in Eq.~(\ref{eq:rate}) grows as $\sim \OO(m_\chi^3)$. As the matrix element (Eq.~(\ref{eq:matrix-element})) needs to be evaluated for each of the considered ions and enters the calculation squared, the overall scaling becomes $\sim \OO(N^2_{\text{atoms}}m_\chi^3)$, where $N_{\text{atoms}}$ is the number of atoms in the primitive cell. Therefore, for systems with large $N_{\text{atoms}}$, considering interactions with heavy DM particles quickly becomes intractable. Therefore, once we reach unfeasible CPU-time requirements, we extrapolate the evaluated reach with a power-law dotted line to guide the eye for large systems in Figures~\ref{fig-dark-photon-sensitivity-DP-crystals} and \ref{fig-operators-sensitivity}.

For the metal-organic framework materials, the phonon bands tend to have a similar density and structure. There is a large density of low-energy ($\lesssim 50$ meV) excitations stemming from all possible torques of the ligands in the mostly empty unit cell, which are separated from the higher-energy displacements, which change the nearest-neighbour distances (Fig.~\ref{fig-band-structures} bottom row). The similarity in the state distribution manifests itself in similar predicted detector sensitivities for all of the considered MOFs in both mediator limits (Fig.~\ref{fig-dark-photon-sensitivity-DP-MOFs}).

This is not the case for all materials in general, as demonstrated in Fig.~\ref{fig-dark-photon-sensitivity-DP-crystals}. The inorganic crystals exhibit a qualitatively different band density and distribution (as shown in Fig.~\ref{fig-band-structures} top row). The broader density of states of LiNbO$_3$ relative to GaAs opens up a bigger phase space for potential transitions, inflating the rate and improving detector sensitivity. LiNbO$_3$ also has a high density of chiral states in the lowest-energy phonon band, increasing its sensitivity (as discussed in Sec.~\ref{sec:dft-and-kappa}). In the all-operator approach (Fig.~\ref{fig-operators-sensitivity}), we can see that the behavior is similar to that of the dark photon model, where the MOFs tend to behave similarly while GaAs underperforms LiNbO$_3$. 

There is a specific sensitivity ordering for the different interaction types where operator $\OO_{1}$ is sensitive to the smallest cross sections as it does not depend (and is therefore not suppressed) by any additional variable. The operators dependent on the momentum transfer $\vect{q}$ ($\OO_{5a}$ and $\OO_{11}$) tend to be disfavored when the mass of the DM candidate is small and therefore does not carry enough momentum to transfer to the target. The operators dependent on the average DM velocity $\vect{v}_\chi$ ($\OO_{5a}$ and $\OO_{8a}$) are then suppressed by this factor, which is of the order $\sim 10^{-3}$. Furthermore, for the operators that do not depend on the mass of the struck fermion, the hadrophilic and leptophilic interaction sensitivities exhibit a similar order of magnitude. Those suppressed by the $\frac{1}{m_\psi}$ factor show a decrease in the hadrophilic interaction rate because of the heavier fermionic target.

As discussed in Sec.~\ref{sec:dft-and-kappa}, we estimate that the probability that an excitation produces a detectable chiral phonon ($\kappa$) is equal to the expectation value of the dimensionless phonon angular momentum in the lowest energy band ($\langle ||\mathbf{J}_1|| \rangle/\hbar$). We then take this parameter and scale the expected rate proportionally for all considered materials. Fig.~\ref{fig-dark-photon-phasespace} shows the sensitivity scaling for several choices of the parameter $\kappa$ for the example of LiNbO$_3$ in the dark photon interaction with a light mediator. We can see that since the assumed number of recorded events scales linearly with $\kappa$, so does the expected sensitivity.

This choice of sensitivity scaling is conservative, since we work under the assumption that our material will host chiral magnetic phonons of a magnitude sufficient to be read out by a surface-based magnetometer at the lowest energy band. Therefore, any excitation at the lowest allowed energy will obey this scaling, while for higher energy transfers, the initial energy deposition will decay into a number of stable low-energy phonons, where only one of them is required to be of sufficient chirality. That is why only the lightest allowed DM candidates will obey this scaling, while heavier particles will be less restricted. In order to calculate this scaling exactly, one would have to calculate the phonon branching ratios at each point in reciprocal space, which is too resource-demanding; therefore we have used this conservative approach.

\section{Photonic backgrounds}\label{sec:background}
The dominant source of background events which are able to mimic a dark matter signal is radioactive impurities embedded in the active material and the surrounding shielding and electronics. These contaminating radioactive isotopes can decay, emitting photons which interact with the surrounding ions, depositing energy and thereby exciting phononic degrees of freedom.

In Ref.~\cite{Berghaus:2021wrp}, Rayleigh scattering was identified as the dominant photon-ion scattering process that is a source of background events for phononic DM detectors. The current well-shielded experiment EDELWEISS-III~\cite{EDELWEISS:2018tde} has been used to estimate this background as an expected photon flux of 0.042 events/kg/day/keV based on the photon--electron Compton scattering rate. 

Such a sizable expected background rate, surpassing that caused by solar neutrinos, implies the necessity of an active veto for detectors reaching kg$\cdot$yr exposures. In addition to the active veto placed around the detector, we can further use the anisotropic distribution of chiral modes in reciprocal space to suppress the photonic background. 

The preference in the direction of momenta of phonons exhibiting a large degree of chirality (see Fig.~\ref{fig-cubes}) leads to a momentum transfer direction that will be favored for chiral phonon readout. As the average momentum deposition follows the flux of the incident DM particles~\cite{Hochberg:2016ntt, Catena:2023awl}, this will lead to a directional response of the detector. We can then use the directionality of the signal to further veto the signal coming from radioactive contamination, as it will not exhibit the daily modulation patterns expected in a DM signal.

\section{Conclusions}
In our previous work~\cite{Romao:2023zqf}, we proposed the use of MOFs for the construction of a chiral phonon quantum sensor for DM direct detection, as well as for other applications, due to its ability to sense $\OO(10\,\text{meV})$ energy depositions. Therein, we focused on one candidate material (InF$_3$(bpy)) and studied its response to DM interactions within the dark photon model. In this work, we expand the explored set of interactions beyond the dark photon model and evaluate all of the non-zero effective interaction operators that are expected in nonmagnetic materials. Furthermore, we study a wide set of MOFs with different geometries and compositions to verify that our initial choice was not an accidental outlier. 

We confirm that the expected sensitivities are similar for all of the studied MOFs and identify the origin of this behavior to a similar distribution of low-energy phononic states in the band structure. This behavior is not universal for all materials, as the two inorganic crystals GaAs and LiNbO$_3$ show a different sensitivity that stems partly from a different phononic state distribution as well as from a different density of chiral states. While the MOFs are overall more sensitive than the inorganic crystals, the inorganic crystals are also projected to surpass current exclusion limits.

An advantage in choosing a noncentrosymmtric MOF as the target material lies in the presence of chiral modes in the low-energy stable bands as well as in the anisotropy of the crystal, leading into an anistotropic momentum distribution of these states. This anisotropy leads to a directional response of the detector that will be able to suppress the photon-ion Rayleigh scattering background that is the dominant source of signal coming from radioactive isotope contamination.

This leads us to the conclusion that once constructing a MOF chiral phonon quantum sensor, the emphasis while choosing a target material should be on the density of the chiral states in the stable band and the magnitude of their magnetic moments since interaction probabilities tend to be similar. As the density of chiral states can be readily calculated \textit{ab initio} using DFT, the major challenge lies in prediction of phonon magnetic moments. While theoretical frameworks describing these effects are emerging \cite{juraschekChiral2025, shabalaAxial2025}, the discovery of new materials with large phonon magnetic moments is, at present, largely driven by experimental approaches.

Our next step in constructing such a sensor is the verification of the ability to read out individual phonons directly through their magnetic response as shown in Fig.~\ref{fig-testing-prototype}, which would also represent a new approach in the search for materials with large phonon magnetic moments. In such a setup we will be able to identify the most promising MOF candidate material for the construction of a new quantum sensor for use in particle physics, photonics, radiation detection and other industrial applications.

\section{Methods}\label{sec:methods}

\textbf{Density functional theory:} DFT and DFPT calculations of Li$_2$(tar), Sr(tar), AgCl(phen), Zn(NH$_4$)(fmt)$_3$, Cd(Gua)(fmt)$_3$, and CuCl(pyr) were performed using the \textsc{Abinit} software package (v. 9) \cite{verstraeteAbinit2025, gonze1997dynamical, bottin2008large, bjorkman2011cif2cell}. The Perdew--Burke--Ernzerhof exchange--correlation functional \cite{perdew1996generalized} was used with the dispersion correction of Grimme \cite{grimme2010consistent}. A plane-wave basis set was used with the default norm-conserving pseudopotentials of the \textsc{Abinit} library. The energy cutoffs used were chosen by convergence studies: 30 Ha (Li$_2$(tar), Sr(tar)), 32 Ha (AgCl(phen), CuCl(pyr)), 34 Ha (Zn(NH$_4$)(fmt)$_3$), and 38 Ha (Cd(Gua)(fmt)$_3$). Monkhorst--Pack grids \cite{monkhorst1976special} of $\mathbf{k}$-points were also chosen by convergence studies. The real-space basis vectors of these grids were: $[-6 ~4 ~4],~[6 ~{-4} ~4],~[6 ~4 ~{-4}]$ (Li$_2$(tar));  $[8 ~0 ~0],~[0 ~8 ~0],~[0 ~0 ~8]$ (Sr(tar), AgCl(phen)); $[3 ~0 ~0],~[0 ~3 ~0],~[0 ~0 ~4]$ (Zn(NH$_4$)(fmt)$_3$); $[4 ~5 ~{-4}],~[4 ~5 ~4],~[0 ~0 ~8]$ (Cd(Gua)(fmt)$_3$); and $[4 ~0 ~0],~[0 ~4 ~0],~[0 ~0 ~4]$ (CuCl(pyr)). Convergence studies were performed using $1\%$ of the internal pressure as the convergence criterion. 

Calculations of the phononic structure were performed on a $2 \times 2 \times 2$ grid of $\mathbf{q}$-points (Sr(tar), AgCl(phen), CuCl(pyr)), or at $\mathbf{q} = (0 ~0 ~0)$ (Li$_2$(tar), Zn(NH$_4$)(fmt)$_3$, Cd(Gua)(fmt)$_3$). The crystal structures were relaxed to an internal pressure of $< 0.1$ GPa within their experimental space group prior to DFPT calculations. These materials are diamagnetic and therefore calculations were performed without spin polarization.  The DFPT-calculated dynamical matrices and Born effective charges of GaAs and LiNbO$_3$ were taken from the literature~\cite{jainCommentary2013, osti_1200591, uedaChiral2025}.

\textbf{Excitation rate calculations:}  The PhonoDark software package (v. 1.1.0) ~\cite{Trickle:2020oki} was used to calculate excitation rates interfaced with Phonopy, an open source package for phonon calculations in Python, with calculations of general effective DM interactions~\cite{Catena:2019gfa}. In order to use results of DFPT calculations from \textsc{Abinit}, the PhonoDark code was modified to accept an input force constant matrix from AbiPy \cite{verstraeteAbinit2025}. A fixed time of day of $t=0$ was chosen for an exposure of kg$\cdot$yr. The exclusion confidence limits were set to 95\% assuming zero background, or equivalently, three recorded events. The DM velocity distribution was modeled with a truncated boosted Maxwell--Boltzmann distribution. 

\textbf{Data availability:} All computational data are publicly available from Ref.~\cite{sdata}.

\acknowledgements{
MM acknowledges the support by the CTU Mobility Project MSCA-F-CZ-III under the number \texttt{CZ.02.01.01/00/22\_010/0008601} and the support of the Ministry of Education, Youth and Sports of the Czech Republic through e-INFRA CZ (ID:90254). FK acknowledges the support of MEYS grant LM2023051 and of the project TERAFIT of the Ministry of Education, Youth and Sports, Czech Republic, co-funded by the European Union (\texttt{CZ.02.01.01/00/22\_008/0004594}). CPR acknowledges support from the project FerrMion of the Ministry of Education, Youth and Sports, Czech Republic, co-funded by the European Union (\texttt{CZ.02.01.01/00/22\_008/0004591}), the European Union and Horizon 2020 through Grant No. 101030352, ETH Zurich, and the Swiss National Supercomputing Center (CSCS) under project IDs s1128 and eth3.}
\bibliography{ref}

\section{Appendix}\label{sec:appendix}

\begin{figure*}[ht]
    \centering \includegraphics[width=0.99\textwidth,trim=3cm 4cm 7cm 0cm,clip]{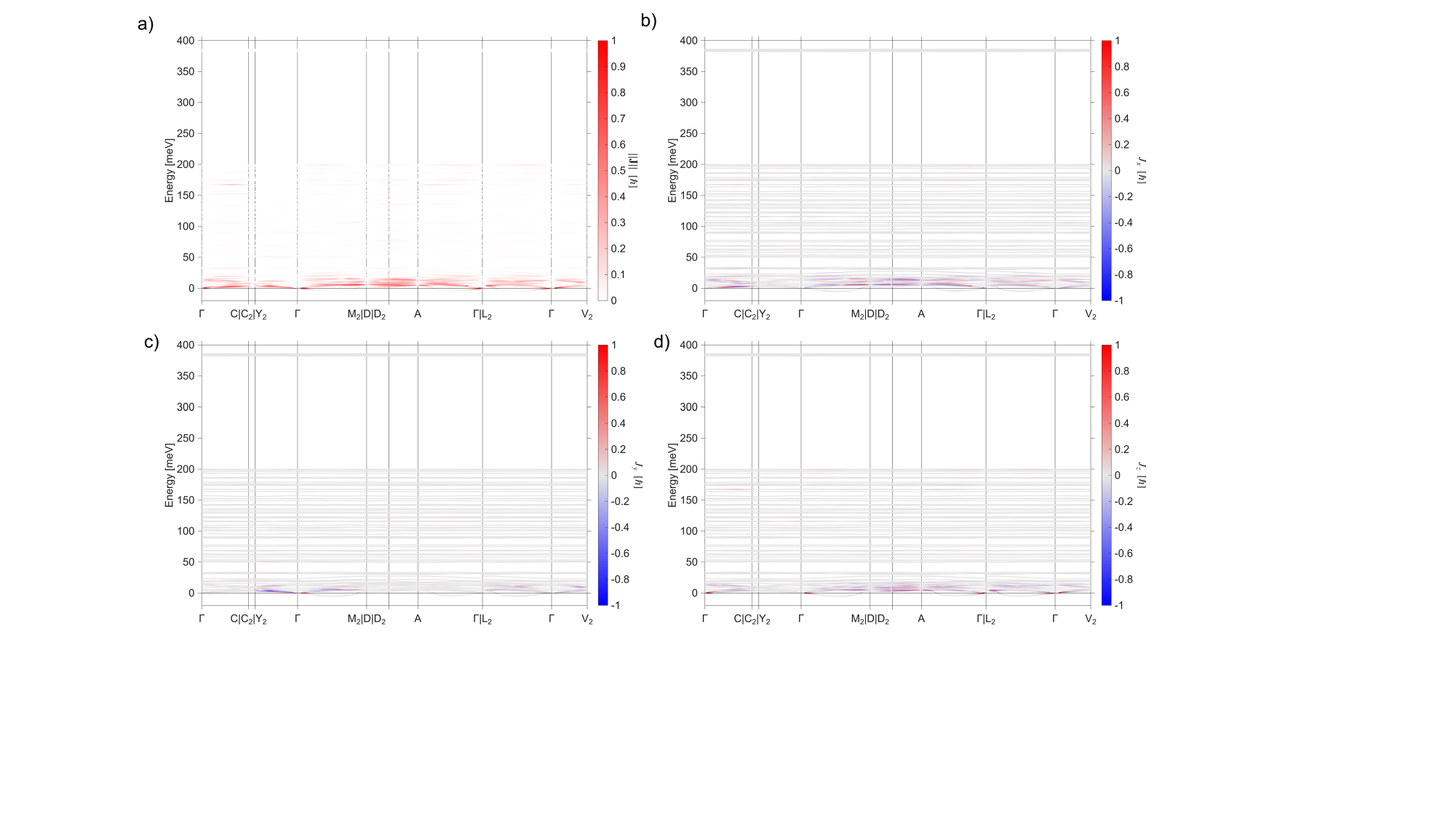}
  \caption{Phonon band structure of AgCl(phen), with bands coloured according to the magnitude (panel \textbf{a}) and  $x$, $y$, and $z$ components of the angular momentum vector $\mathbf{J}$ (panels \textbf{b}-\textbf{d}). Special points in and paths through the Brillouin zone were chosen following the literature \cite{hinuma2017band}.}
  \label{fig-AgClphenJ}
\end{figure*}

\begin{figure*}[ht]
    \centering    \includegraphics[width=0.99\textwidth,trim=3cm 4cm 7cm 0cm,clip]{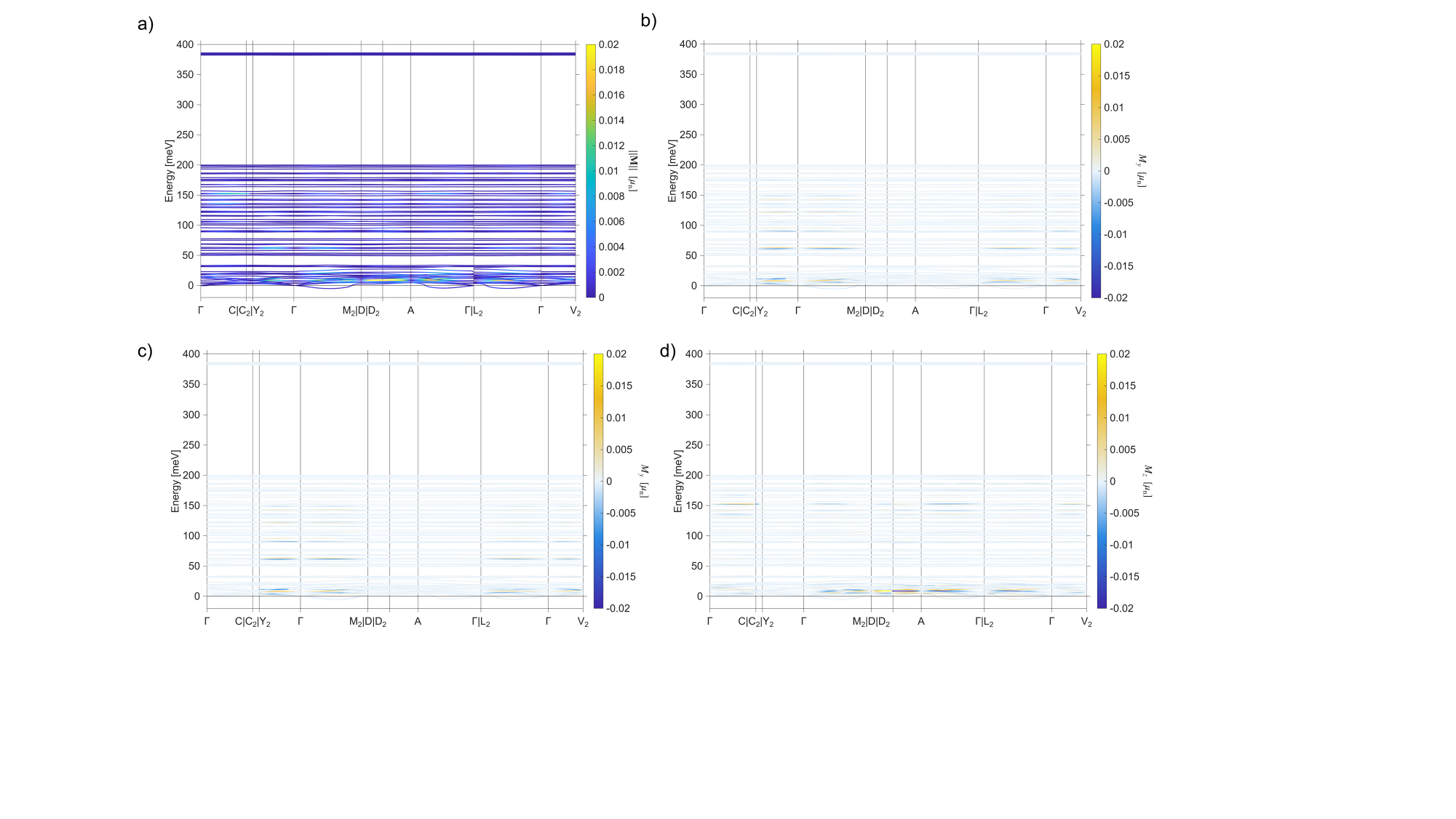}
  \caption{Phonon band structure of AgCl(phen), with bands coloured according to the magnitude (panel \textbf{a}) and  $x$, $y$, and $z$ components of the magnetic moment vector $\mathbf{M}$ (panels \textbf{b}-\textbf{d}), in units of the nuclear magneton; as calculated from the motion of the Born effective charges \cite{juraschek2019orbital, chaudharyAnomalous2025}. Special points in and paths through the Brillouin zone were chosen following the literature \cite{hinuma2017band}.}
  \label{fig-AgClphenM}
\end{figure*}

\begin{figure*}[ht]
    \centering \includegraphics[width=0.99\textwidth,trim=3cm 2.5cm 6cm 0cm,clip]{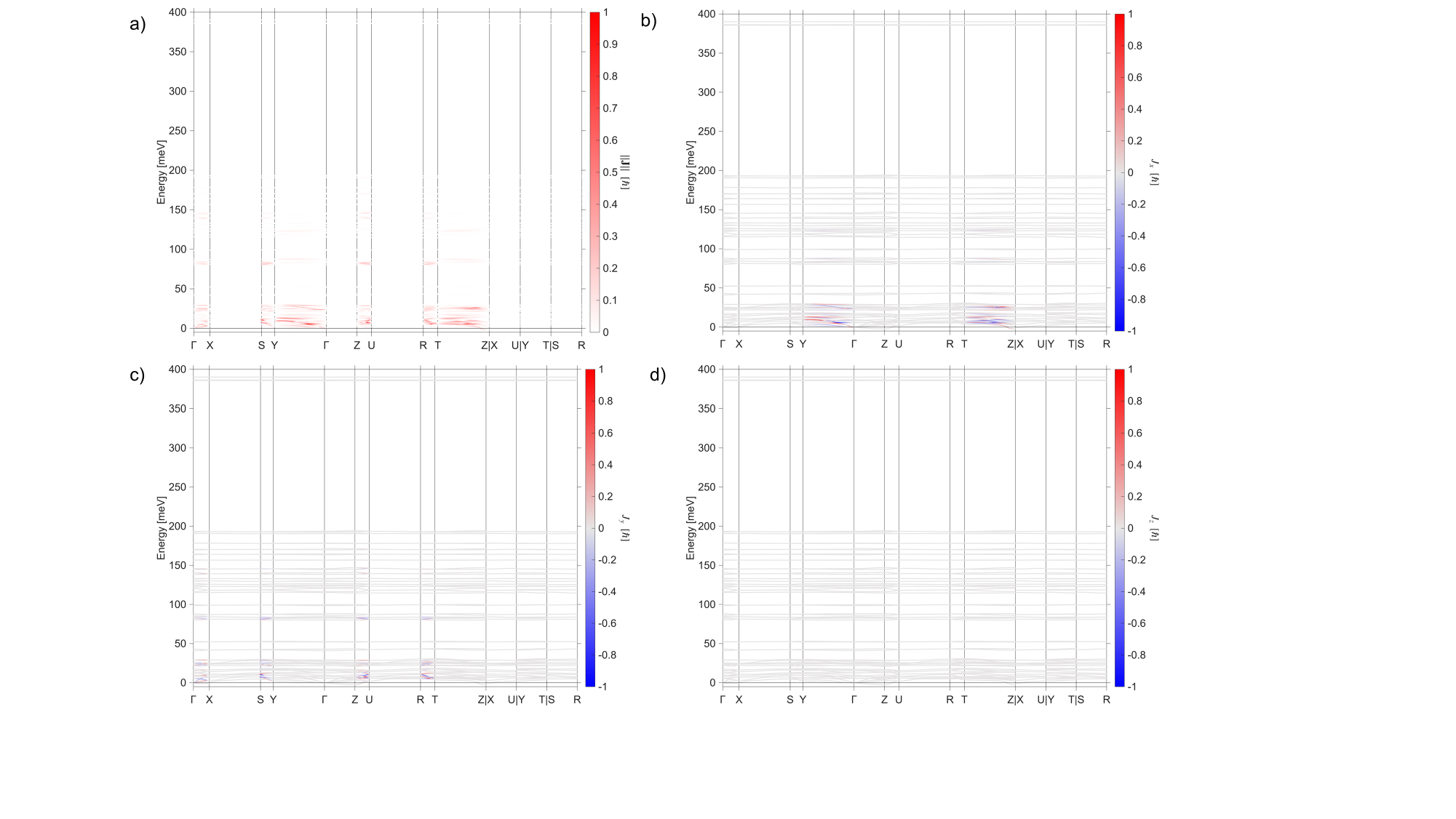}
  \caption{Phonon band structure of CuCl(pyr), with bands coloured according to the magnitude (panel \textbf{a}) and $x$, $y$, and $z$ components of the angular momentum vector $\mathbf{J}$ (panels \textbf{b}-\textbf{d}). Special points in and paths through the Brillouin zone were chosen following the literature \cite{hinuma2017band}.}
  \label{fig-CuClpyrJ}
\end{figure*}

\begin{figure*}[ht]
    \centering    \includegraphics[width=0.99\textwidth,trim=3cm 3cm 6cm 0cm,clip]{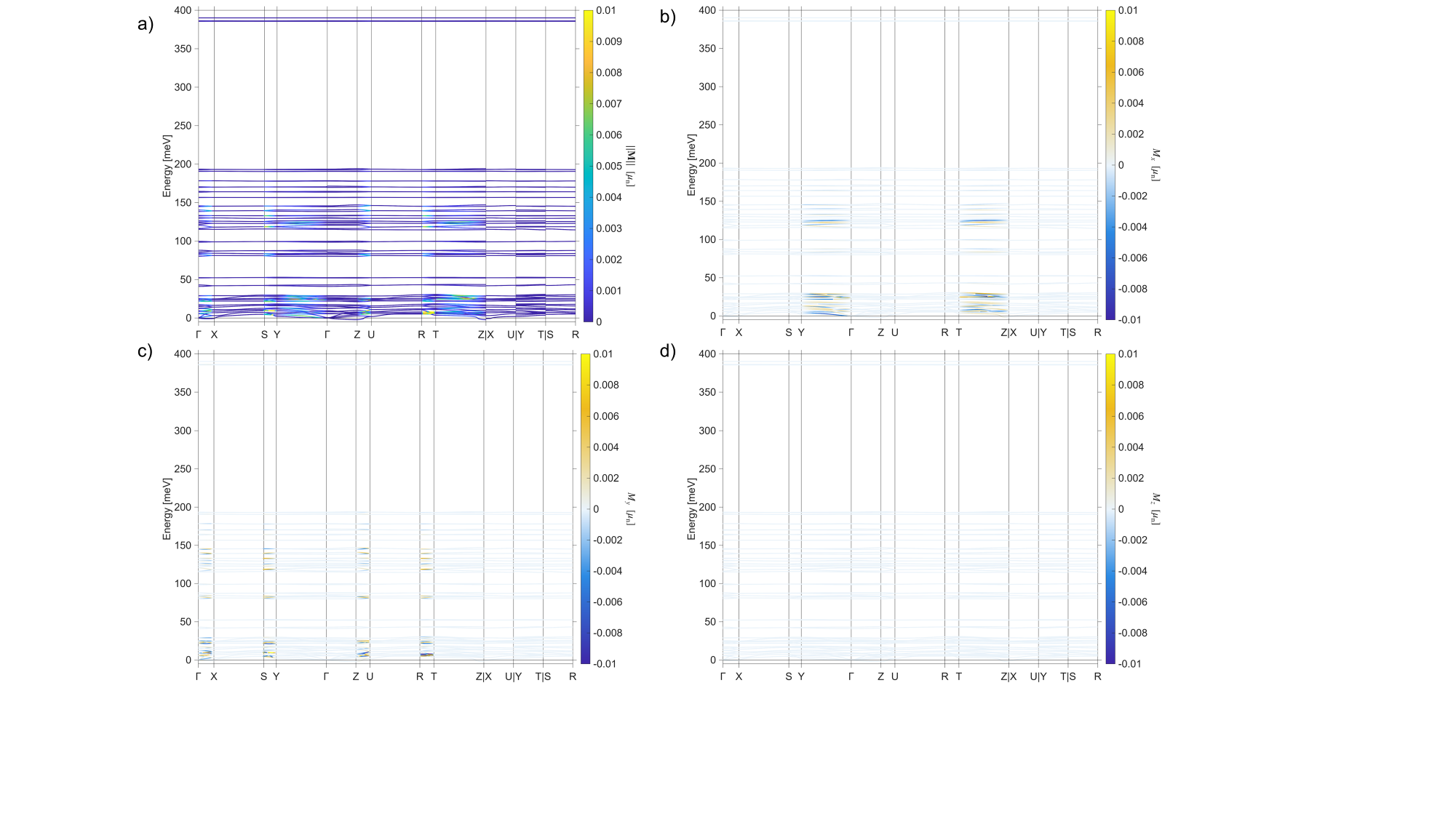}
  \caption{Phonon band structure of CuCl(pyr), with bands coloured according to the magnitude (panel \textbf{a}) and  $x$, $y$, and $z$ components of the magnetic moment vector $\mathbf{M}$ (panels \textbf{b}-\textbf{d}), in units of the nuclear magneton; as calculated from the motion of the Born effective charges \cite{juraschek2019orbital, chaudharyAnomalous2025}. Special points in and paths through the Brillouin zone were chosen following the literature \cite{hinuma2017band}.}
  \label{fig-CuClpyrM}
\end{figure*}

\begin{figure*}[ht]
    \centering \includegraphics[width=0.99\textwidth,trim=3cm 2.5cm 7cm 0cm,clip]{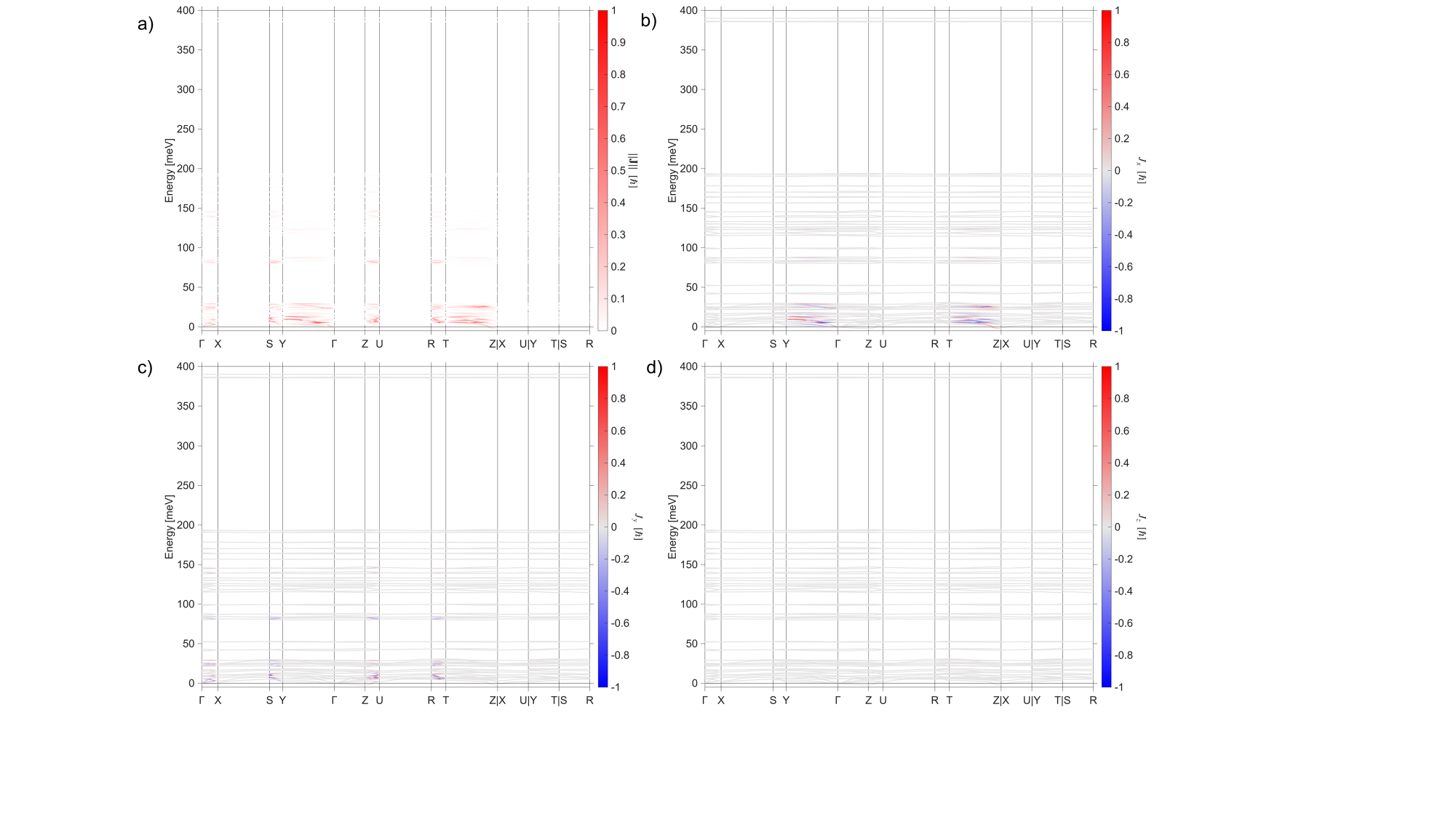}
  \caption{Phonon band structure of Sr(tar), with bands coloured according to the magnitude (panel \textbf{a}) and $x$, $y$, and $z$ components of the angular momentum vector $\mathbf{J}$ (panels \textbf{b}-\textbf{d}). Special points in and paths through the Brillouin zone were chosen following the literature \cite{hinuma2017band}.}
  \label{fig-SrtarJ}
\end{figure*}

\begin{figure*}[ht]
    \centering    \includegraphics[width=0.99\textwidth,trim=3cm 3cm 7cm 0cm,clip]{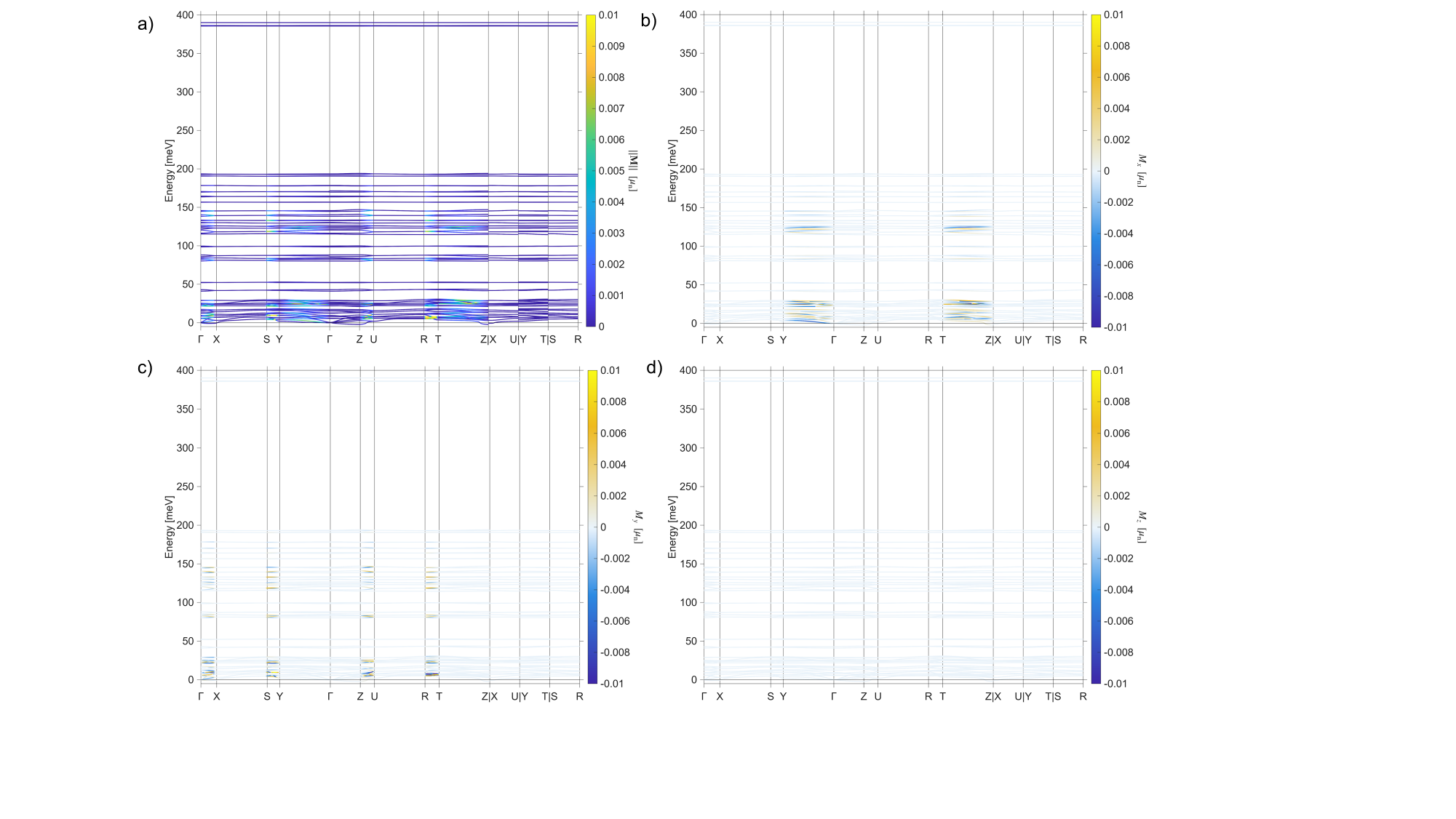}
  \caption{Phonon band structure of Sr(tar), with bands coloured according to the magnitude (panel \textbf{a}) and  $x$, $y$, and $z$ components of the magnetic moment vector $\mathbf{M}$ (panels \textbf{b}-\textbf{d}), in units of the nuclear magneton; as calculated from the motion of the Born effective charges \cite{juraschek2019orbital, chaudharyAnomalous2025}. Special points in and paths through the Brillouin zone were chosen following the literature \cite{hinuma2017band}.}
  \label{fig-SrtarM}
\end{figure*}

\begin{figure*}[ht]
    \centering \includegraphics[width=0.99\textwidth,trim=3cm 2cm 7cm 0cm,clip]{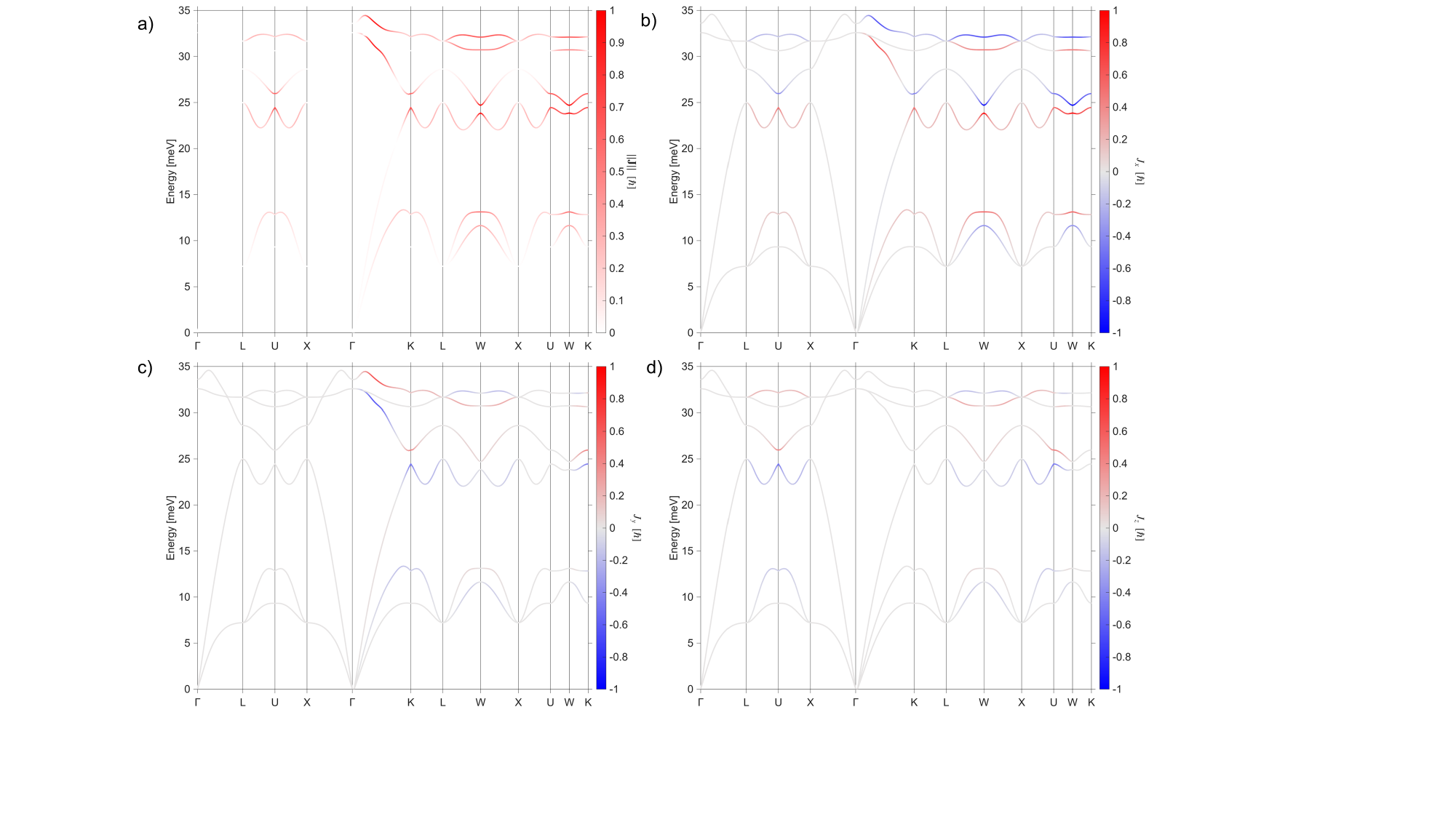}
  \caption{Phonon band structure of GaAs, with bands coloured according to the magnitude of their phonon angular momentum (panel \textbf{a}), and the $x$, $y$, and $z$ components of the angular momentum vector $\mathbf{J}$ (panels \textbf{b}-\textbf{d}). Special points in and paths through the Brillouin zone were chosen following the literature \cite{hinuma2017band}. Dynamical matrices and Born effective charges were obtained from the literature~\cite{osti_1200591}.}
  \label{fig-GaAsJ}
\end{figure*}

\begin{figure*}[ht]
    \centering    \includegraphics[width=0.99\textwidth,trim=3cm 2cm 7cm 0cm,clip]{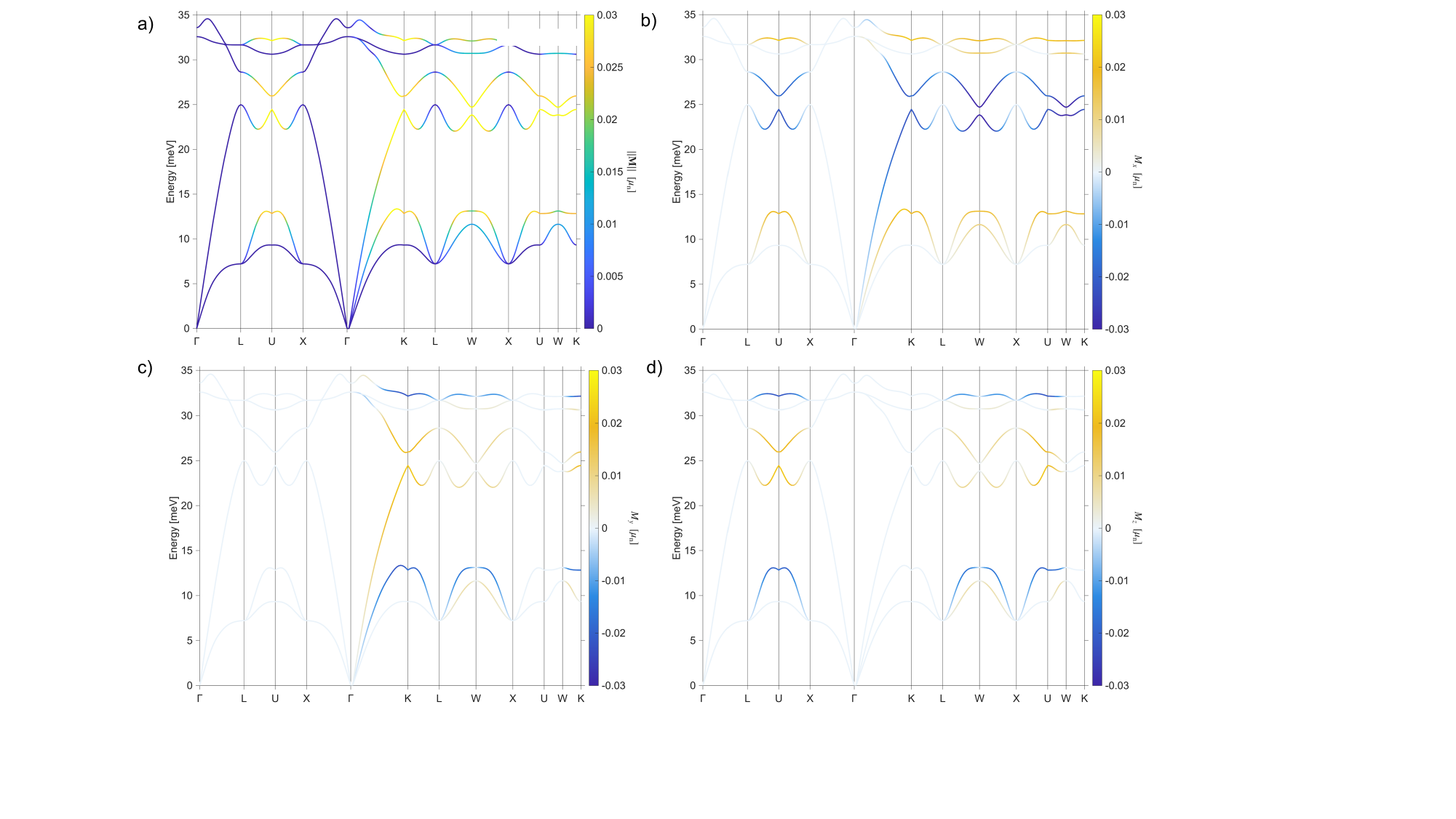}
  \caption{Phonon band structure of GaAs, with bands coloured according to the magnitude (panel \textbf{a}) and  $x$, $y$, and $z$ components of the magnetic moment vector $\mathbf{M}$ (panels \textbf{b}-\textbf{d}), in units of the nuclear magneton; as calculated from the motion of the Born effective charges \cite{juraschek2019orbital, chaudharyAnomalous2025}. Special points in and paths through the Brillouin zone were chosen following the literature \cite{hinuma2017band}. Dynamical matrices and Born effective charges were obtained from the literature~\cite{osti_1200591}.}
  \label{fig-GaAsM}
\end{figure*}

In this Appendix we present the phonon band structures of AgCl(phen), CuCl(pyr), Sr(tar), and GaAs, with bands coloured by the magnitudes and Cartesian components of the phonon angular momentum vector ($\mathbf{J}$), and the phonon magnetic moment vector ($\mathbf{M}$) (Figs.~\ref{fig-AgClphenJ}--\ref{fig-GaAsM}), as calculated using Eqs.~(\ref{eq-pam}) and~(\ref{eq-magmo}). Similar plots for LiNbO$_3$ can be found in Ref.~\cite{uedaChiral2025}. For the calculation of the phonon magnetic moment, we include only the contributions of the motion of the Born effective charges~\cite{juraschek2017dynamical,juraschek2019orbital}, which typically yields magnetic moments less than the nuclear magneton ($\mu_\mathrm{n}$), orders of magnitude smaller than the moments on the order of the Bohr magneton which have been observed in various experiments~\cite{shabalaAxial2025}. However, such calculations can be useful to identify cases where the circular polarizations of different sublattices have an antiferroic ordering, leading to divergent gyromagnetic ratios, where magnetism arises from modes with vanishing total phonon angular momentum ~\cite{chaudharyAnomalous2025}. 

\end{document}